\documentclass[11pt,showpacs,amsmath,amssymb,aps,floats,aps,floatfix,nofootinbib]{article}%

\usepackage{jheppub}
\usepackage{mathtools}
\usepackage{epsfig}
\usepackage{dcolumn}
\usepackage{bm}
\usepackage{amsthm}
\usepackage{amsmath}
\usepackage{amsfonts}
\usepackage{amssymb}
\usepackage{graphicx}
\usepackage{latexsym}
\usepackage{rotating}
\usepackage{multirow}
\usepackage{slashed}
\usepackage{color}
\usepackage{hyperref}
\usepackage{graphicx}%
\usepackage{diagbox}
\usepackage{cancel}
\usepackage{paralist}
\usepackage{tcolorbox}
\usepackage{color, colortbl}
\usepackage{soul}
\usepackage{microtype}

\usepackage{makecell}
\usepackage{collcell}
\newcolumntype{G}{>{\collectcell\gape}{c}<{\endcollectcell}}
\newcolumntype{A}{>{\collectcell\gape}{l}<{\endcollectcell}}
\newcolumntype{Z}{>{\collectcell\gape}{r}<{\endcollectcell}}

\usepackage{amsmath,amsfonts,bm}

\def\eqref#1{equation~\ref{#1}}

\def\1{\bm{1}}

\def\rvb{{\mathbf{b}}}

\def\rvg{{\mathbf{g}}}
\def\rvh{{\mathbf{h}}}

\def\rvx{{\mathbf{x}}}

\def\rmK{{\mathbf{K}}}

\def\rmW{{\mathbf{W}}}

\DeclareMathAlphabet{\mathsfit}{\encodingdefault}{\sfdefault}{m}{sl}
\SetMathAlphabet{\mathsfit}{bold}{\encodingdefault}{\sfdefault}{bx}{n}

\newcommand{\Ls}{\mathcal{L}}

\definecolor{Gray}{gray}{0.95}

\theoremstyle{definition}

\providecommand\inspire[1]{\href{https://inspirehep.net/search?p=find+#1}{{\tiny IN}{\footnotesize SPIRE}}}

\providecommand{\jhep}[3] {\ifnum#2>2009%
\href{https://doi.org/10.1007/JHEP#1(#2)#3}{\emph{JHEP} {\bf #1} (#2) #3}%
\else%
\href{https://doi.org/10.1088/1126-6708/#2/#1/#3}{\emph{JHEP} {\bf #1} (#2) #3}%
\fi}%

\def\issueFromCounter.#1#2#3#4#5#6.{#2#3}
\providecommand{\jstat}[2]{\PackageWarningNoLine{\jname}{The macro \protect\jstat\space is obsolete!\MessageBreak Please typeset JSTAT as any other journal}%
  \href{https://doi.org/10.1088/1742-5468/#1/\issueFromCounter.#2./#2}{\emph{J.\ Stat.\ Mech.\ }(#1) #2}}

\providecommand{\arXivid}[1]{\href{https://arxiv.org/abs/#1}{\tt arXiv:#1}}
\providecommand{\Math}[2]{%
\if!#1!%
\href{https://arxiv.org/abs/math/#2}{\tt math/#2}%
\else%
\href{https://arxiv.org/abs/math.#1/#2}{\tt math.#1/#2}%
\fi}

\newcommand{\pt}{p_{\rm T}}
\newcommand{\W}{$W$}
\newcommand{\GeV}{{\rm GeV}}
\newcommand{\MD}{\texttt{MD}}

\newcommand{\gist}[1]{}  
\newcommand{\mynote}[1]{}

\title{Invariant representation driven neural classifier for anti-QCD jet tagging}

\author{Taoli~Cheng}
\author{and Aaron~Courville}

\affiliation{Mila --- Quebec Artificial Intelligence Institute,\\
St-Urbain, Montreal, QC H2S 3H1, Canada}
\affiliation{Department of Informatics, Universit\'{e} de Montr\'{e}al,\\
Chemin de la Tour, Montreal, QC H3T 1J4, Canada}
\emailAdd{chengtaoli.1990@gmail.com}
\emailAdd{aaron.courville@umontreal.ca}

\abstract{We leverage representation learning and the inductive bias in neural-net-based Standard
  Model jet classification tasks, to detect non-QCD signal jets. In establishing the framework for
  classification-based anomaly detection in jet physics, we demonstrate that, with a
  \emph{well-calibrated} and \emph{powerful enough feature extractor}, a well-trained
  \emph{mass-decorrelated} supervised Standard Model neural jet classifier can serve as a strong
  generic anti-QCD jet tagger for effectively reducing the QCD background. Imposing
  \emph{data-augmented} mass-invariance (and thus decoupling the dominant factor) not only
  facilitates background estimation, but also induces more substructure-aware representation
  learning. We are able to reach excellent tagging efficiencies for all the test signals
  considered. In the best case, we reach a background rejection rate of 51 and a significance
  improvement factor of 3.6 at 50\% signal acceptance, with the jet mass decorrelated. This study
  indicates that supervised Standard Model jet classifiers have great potential in general new
  physics~searches.}

\begin{document}
\maketitle\flushbottom

\section{Introduction}

 Data-driven beyond Standard Model~(SM) new physics searches at the Large Hadron Collider~(LHC) have been explored in the regime of autoencoders~\cite{Heimel-ml-2018mkt, Farina-ml-2018fyg, Blance-ml-2019ibf, Ostdiek-ml-2021bem, Finke-ml-2021sdf, Hajer-ml-2018kqm, Dillon-ml-2021nxw}, latent variable models~\cite{Cerri-ml-2018anq, Cheng-ml-2020dal}, density estimation~\cite{Nachman-ml-2020lpy, Hallin-ml-2021wme, Stein-ml-2020rou}, and weakly supervised classification~\cite{Collins-ml-2018epr, ATLAS-ml-2020iwa}.
At the same time, a community-wide challenge~\cite{Kasieczka-ml-2021xcg} for simulated signal detection at the LHC turned out to be short of effective enough solutions (for identifying all the test signals) despite the efforts in exploring different generative models, including variational autoencoders (and variants) and flow-based models.
Even though the methods might be augmented with other strategies such as semi-supervision, the backbones of many of them are constrained to the framework of the generative approach.\footnote{We introduce the terminologies ``discriminative'' and ``generative'', in the current context, to express the fundamental differences between classification-based supervised learning and data likelihood-driven unsupervised learning. Especially, when we talk about ``generative'' modelling, we mean the underlying modelling for the joint distribution $p(\rvx, y)$ of the data and labels rather than the generative ability. In contrast, ``discriminative'' approaches often directly model the posterior probability~$p(y | \rvx)$.}
Relatively speaking, generative modelling has the advantage of not relying on signal examples or labelled simulation data. However, it has been observed that even perfect density estimation can't guarantee effective out-of-distribution (OoD) detection~\cite{bib2018arXiv181009136N,DBLP-ml-journals-ml-corr-ml-abs-1812-04606, DBLP-ml-journals-ml-corr-ml-abs-2012-03808}. Similar phenomena have been observed in generative model-based anomalous jet tagging~\cite{Cheng-ml-2020dal} and later studied in the setting of autoencoders~\cite{Finke-ml-2021sdf}. Variational Autoencoders trained on background QCD jets might assign even higher likelihoods to non-QCD signal jets, casting doubt on the robustness of these approaches. In contrast to these generative model-based approaches, classifier-based anomalous jet tagging is largely~unexplored.

On the other hand, discriminative classifier-based out-of-distribution detection~\cite{DBLP-ml-journals-ml-corr-ml-HendrycksG16c, DBLP-ml-journals-ml-corr-ml-abs-1812-04606, bib2016arXiv161201474L, bib2018arXiv180204865D, bib2018arXiv180210501M, DBLP-ml-journals-ml-corr-ml-abs-1908-05569, DBLP-ml-journals-ml-corr-ml-abs-2003-02037, DBLP-ml-journals-ml-corr-ml-abs-2006-10108, DBLP-ml-journals-ml-corr-ml-abs-2007-05134}, as an alternative approach, has been well-studied in the machine learning community. It has a few advantages especially due to the outstanding capacity of deep neural networks. By incorporating task-aware inductive biases within the classifier, it increases out-of-distribution awareness and sensitivity compared with the generative counterpart. A discriminative classifier is also equipped to learn useful invariances, which can be leveraged to detect novel patterns, from the data.
At the same time, it is shown that, in jet physics, the classification-induced latent representations could be meaningful and generalizable across similar tasks~\cite{Cheng-ml-2019isq}. Combining these threads (representation learning, cross-task transferability, and anomaly detection), we explore the potential of classification-induced representations in searching for new physics signals, and establish the framework of model training, inference, and evaluation.

A few studies have been carried out in the regime of classification.~\cite{Aguilar-Saavedra-ml-2017rzt, Aguilar-Saavedra-ml-2021utu} design a dedicated QCD/non-QCD classifier for detecting potential signals, where the anti-QCD class consists of different signal decay modes. The classifier is built on high-level features, which don't fully reveal the powerful representation learning of deep neural networks.~\cite{khosa2020anomaly} utilizes a Standard Model jet classifier to detect new physics signals, with an auxiliary task of outlier exposure.
These previous studies either design or inject some information about the target signals.
And it was believed, in the community, that target Standard Model jet classifiers (e.g., W or Top tagger) are not able to serve as generic anomaly detectors. This belief has also driven many unsupervised learning approaches for anomalous jet tagging recently.
A systematic study in exploring the OoD detection ability of SM jet classifiers is still lacking.
However, we show that the SM neural jet classifiers can serve as very effective anti-QCD taggers due to the rich capacity in representation learning.
While taking in low-level features, it's possible to induce generalizable representations, which can be used to effectively detect out-of-distribution non-Standard-Model signals.
This paradigm shift, in which we switch from designing dedicated non-QCD taggers to directly utilizing existing Standard Model jet classifiers to detect unseen signals, ultimately leads to a unified path of joint efforts from two historically separate research threads: jet classifier architectures and anomaly detection applications.

More concretely, in the context of heavy resonance searches, we explore the limit of utilizing low-level information-equipped neural jet classifiers for anti-QCD jet tagging. We train the model to classify Standard Model jets, but re-utilize the model as a generic anomalous jet tagger.
Slightly different from the usual OoD detection in machine learning, where one treats all the in-distribution classes equally, here we leverage the classification as class-conditional OoD detection (i.e., QCD vs non-QCD classification in the anomaly detection phase).
The main building blocks include in-distribution (InD) training classes, representation extractors (physics-oriented jet classifiers), and anomaly detection scenarios (anomaly score, OoD tagging strategy):
\begin{itemize}
    \item \textbf{Datasets.} A well-designed and inclusive set of in-distribution classes fully leverages the representational power of deep neural nets and thus facilitates better OoD detection performance. Including all the boosted Standard Model jets within the LHC kinematic range equivalently injects the domain knowledge on particle types and underlying interactions, and expects the best tagging efficiency.
    \item \textbf{Neural architecture.} On the one hand, modern, high-capacity neural architectures are crucial for effective classification and representation extraction. On the other hand, physics-friendly architectures induce better generalization ability, which is also an important ingredient for effective OoD detection. There are great efforts in the community to design sophisticated  physics-motivated architectures. These efforts can be re-utilized for generic new physics searches.
        \item \textbf{OoD identifier.} When the classifier is not directly trained with / exposed to OoD examples, we normally utilize post hoc OoD identifiers interfaced with the trained model to select anomalous events. The learned representations can be interfaced to an OoD identifier, either directly through the latent vectors, or via the classification predictions.
    \item \textbf{Predictive uncertainty estimates.} The calibration and uncertainty estimates of neural classifiers~\cite{ pmlr-v70-guo17a, Minderer2021RevisitingTC} provide a better guarantee for trustworthy real-world deployment of deep neural networks. Methods improving uncertainty estimates have been shown to help with OoD detection as well.
            \end{itemize}

Aside from these common aspects in OoD detection, there is a special issue in anomalous jet tagging.
In model-independent heavy resonance searches, decorrelating the jet mass from a tagger~\cite{Dolen-ml-2016kst, ATL-PHYS-PUB-2018-014, Bradshaw-ml-2019ipy} is desired to facilitate effective background estimation. In supervised jet tagging, mass decorrelation generally decreases tagging performance. However, from a different perspective, we show that a mass-decorrelated supervised jet classifier can serve as a generic anomalous jet tagger of outstanding performance. \emph{Imposing mass-invariance on the classification task turns out to enhance (subdominant) feature learning effectively.}

We employ boosted SM jets (QCD/W/Top) as in-distribution classes and utilize the learned representations to perform anti-QCD jet tagging. We test on jets originating from hypothesized new physics heavy resonances. Regarding strategies to improve uncertainty estimate, we employ methods including deep ensembling, one-vs-all and all-vs-all combinational softmax probability, and distance-preserving Gaussian Process. At inference time, to identify non-QCD jets, we use anomaly scoring functions based either on the softmax probability of being non-QCD or on the latent distance to QCD jets. Despite the fact that supervised neural classifiers are designed for specific classification tasks, we demonstrate that a mass-decorrelated SM jet classifier can act as an effective generic anti-QCD tagger.

This paper is organized as follows: in section~\ref{sec:methods}, we depict the schematic and present the methodology. We present a QCD/W classifier viewed as an anomaly detector and reveal the potential of a mass-decorrelated jet classifier serving as a generic anti-QCD tagger.
The experimental setup (including the datasets, the neural architecture, and the training settings) is introduced in section~\ref{sec:setup}. Anomaly detection performance without mass-decorrelation is recorded and analyzed in section~\ref{sec:results}. Then the mass correlation effects and mass-decorrelated anti-QCD tagger are investigated.
In section~\ref{sec:discussion}, we comparatively discuss anomalous jet tagging thematics in discriminative models vs generative models. Finally, in section~\ref{sec:summary}, we summarize this work and present the conclusions.

\section{Methodology}
\label{sec:methods}

\subsection{Supervised classifier in anomaly detection: an experiment}
\label{subsec:exp}

As an illustrative experiment, we first present a simple Standard Model jet classifier serving as an anomalous jet tagger (experimental details can be found in appendix~\ref{app:arch} and section~\ref{sec:setup}). As shown in table~\ref{tab:qcdw}, an accurate QCD/W jet classifier reaches an Area Under the Receiver Operating Characteristic Curve (AUC) of 0.979.
If we interpret the softmax probability $p_{\rm QCD} = p(y=0|\rvx)$ as the probability of a jet being QCD-like, we can leverage the QCD/W neural classifier as an anti-QCD jet tagger.
However, if we directly identify jets with a high non-QCD score $-p_{\rm QCD}$ (equivalently $p_{W}$ since for a QCD/W classifier we have $p_{\rm QCD} + p_{W} = 1$) as inclusive non-QCD jets, discriminating top jets against the QCD background gives an AUC of 0.783.

Surprisingly, for the mass-decorrelated\footnote{The mass decorrelation is achieved by data augmentation in the mass dimension. We mass-augment and resample the W jets to match the mass distributions of QCD and W\@. This procedure will be discussed in section~\ref{sec:setup} and~\ref{sec:results}.} QCD/W classifier, we see promising anti-QCD tagging performance of improved tagging efficiency. (This observation also applies to a QCD/Top classifier.) Though the data are generated under different settings, we find that our mass-decorrelated QCD/W classifier gets an AUC of 0.861 for discriminating between the QCD and top samples from ref.~\cite{Heimel-ml-2018mkt} which studies the autoencoder-based anti-QCD tagging strategy. This is encouraging compared with the AUCs reported there (ranging from 0.63 to 0.78 depending on the decorrelation strength). Remembering that the result is already under distribution shift, we expect even improved performance if the model is trained on data with the same setting as in ref.~\cite{Heimel-ml-2018mkt}.
In comparison, the un-decorrelated classifier displaying a predominant correlation with the jet mass will strongly identify a jet with a mass close to that of the in-distribution W jets (80\,GeV) as non-QCD jets, while suppressing other discriminative factors.

This experiment shows great potential in utilizing the supervised jet classifier as a generic anti-QCD jet tagger. In the following, we will explore the limit of neural jet classifiers performing in OoD detection.

\begin{table}
    \centering
\renewcommand{\arraystretch}{1.2}    \begin{tabular}{cc|cc}
    \hline
    Model     & W & Top & Top from~\cite{Heimel-ml-2018mkt} \\ \hline
    QCD/W  & 0.979 & 0.783 & 0.776 \\
    QCD/W ($\cancel{M}$) & 0.958 & 0.860 & 0.861 \\
    \hline
    \end{tabular}
    \quad
    \begin{tabular}{cc|c}
    \hline
    Model     & Top & W \\ \hline
    QCD/Top    & 0.966  &  0.891 \\
    QCD/Top ($\cancel{M}$)  & 0.876 & 0.915\\
    \hline
    \end{tabular}
    \caption{OoD detection performance in AUC for a (mass-decorrelated ---  $\cancel{M}$) binary jet classifier with $-p_{\rm QCD}$ as the anomaly score. \emph{Left}: QCD/W classifier tested on Top; \emph{Right}: QCD/Top classifier tested on~W.\label{tab:qcdw}}
\end{table}

\subsection{Classifier-based anomaly identifier: general approach}

\begin{figure}
    \centering
    \includegraphics[width=0.7\textwidth]{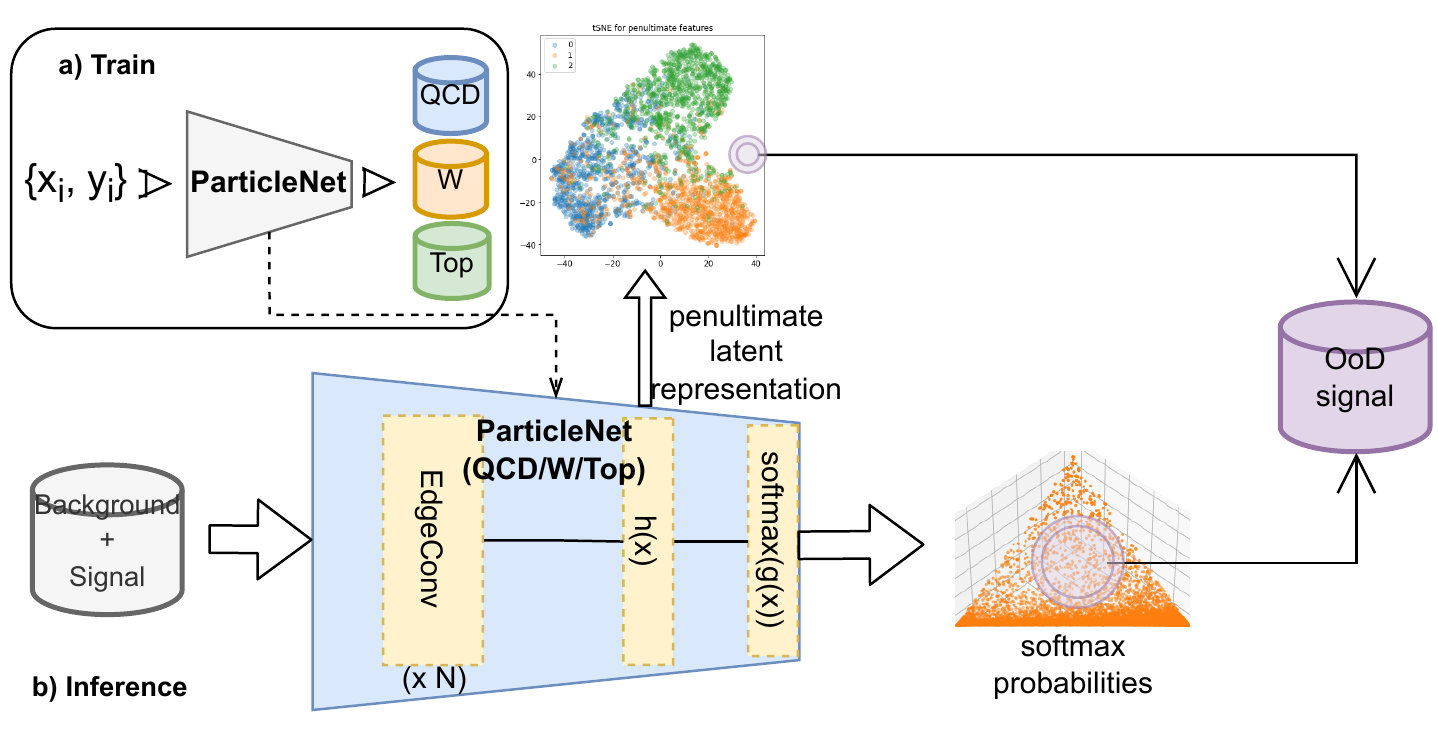}
    \caption{The schematic of neural classifier based anti-QCD jet tagging. The workflow consists of two steps. In the train step (a), we choose the \emph{in-distribution classes} and the \emph{feature extractor}. In the inference step (b), we design post hoc \emph{anomaly scoring functions} operating on the trained classifier model for signal~detection.\label{fig:clfad-schematic}}
\end{figure}

The general approach of employing a supervised classifier as an anomalous jet tagger is depicted in the schematic in figure~\ref{fig:clfad-schematic}. First we train a QCD/W/Top 3-class classifier simply on all the in-distribution SM classes $\{(\rvx_i, y_i); i=1,\dots,N\}$. When we employ the QCD/W/Top 3-way classifier as an anomalous jet tagger, we take advantage of the subtle substructures, which are critical for discriminating a rich set of SM jets, thus possibly generalizable to identifying new signals.
We utilize a powerful feature extractor\footnote{In this study, we employ a Graph Neural Net called ParticleNet. (Check section~\ref{sec:setup} and appendix~\ref{app:arch} for~details.)} for effective representation learning ($\rvx \mapsto \rvh(\rvx)$). Post hoc anomaly scoring functions, depending on the trained classifier, are used for identifying non-QCD jets in the inference~phase.

\begin{figure}
     \centering
     \includegraphics[width=0.32\textwidth]{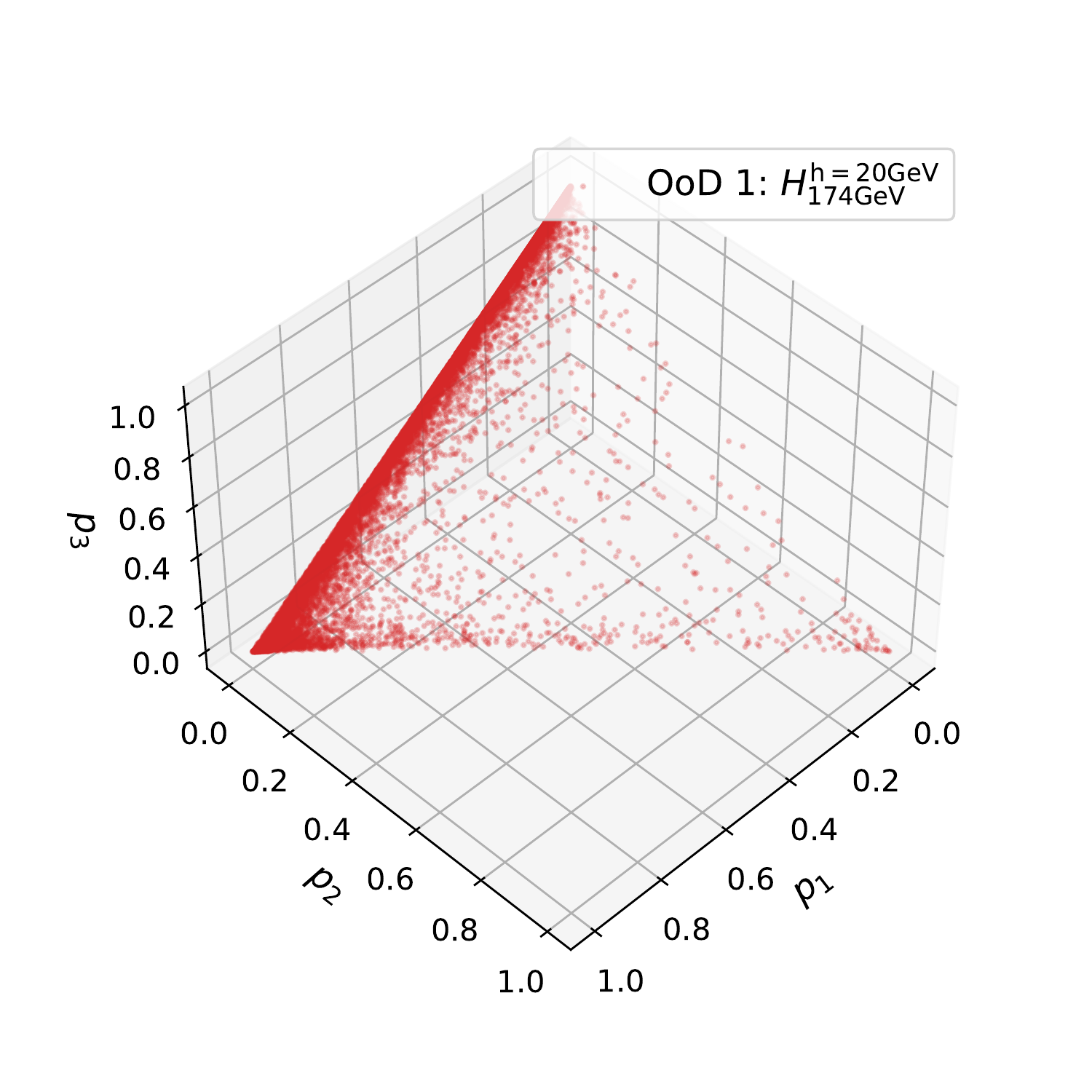}
     \includegraphics[width=0.32\textwidth]{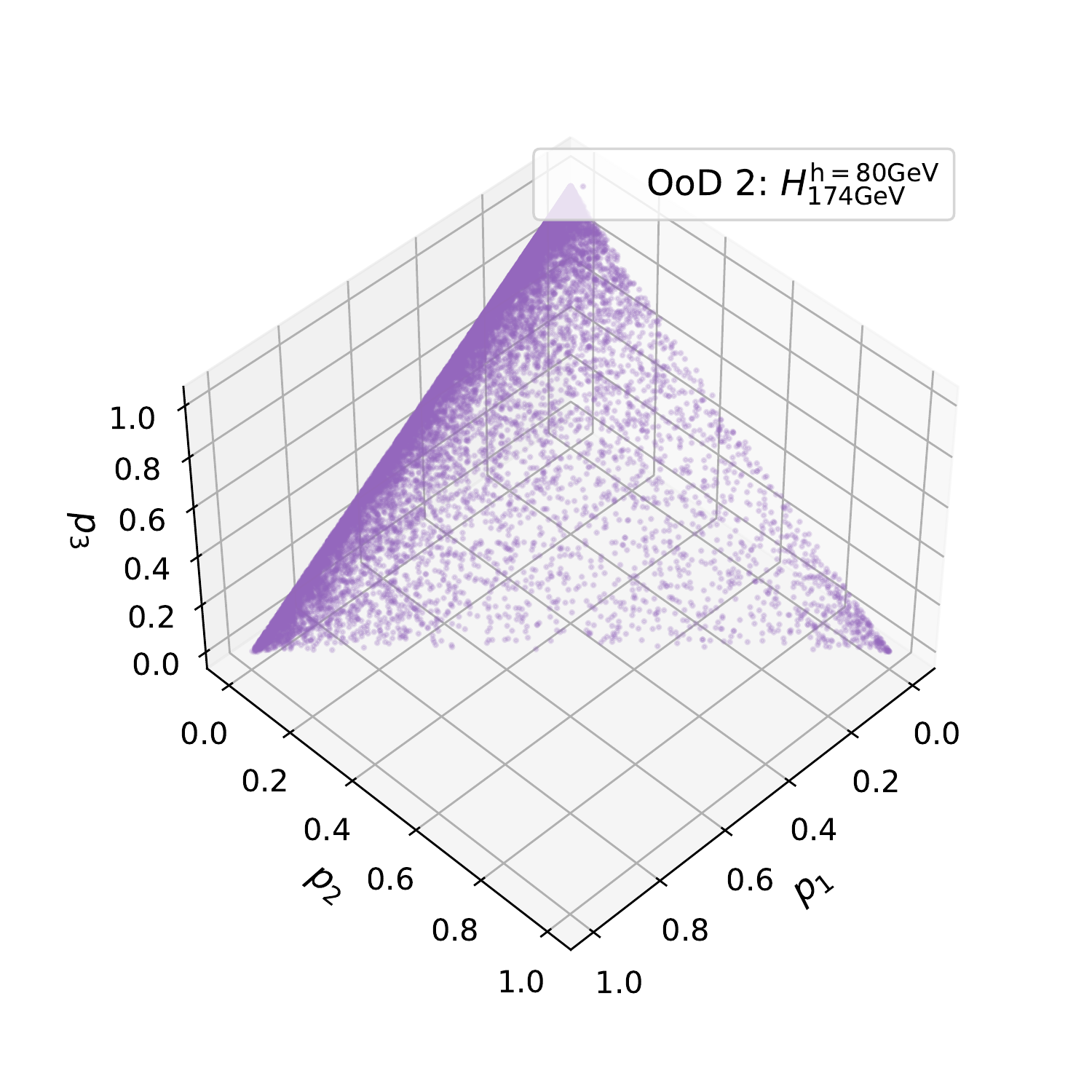}
     \includegraphics[width=0.32\textwidth]{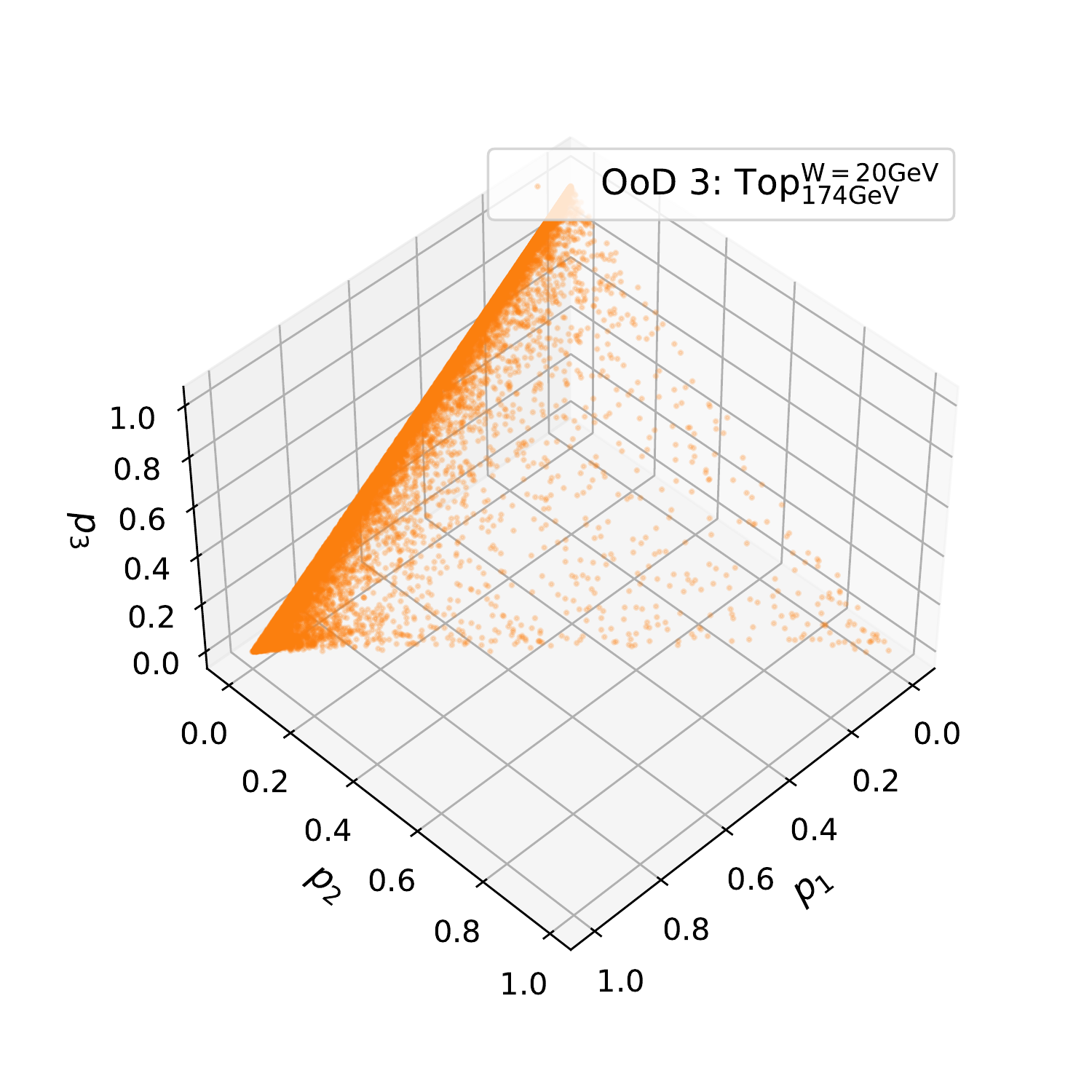}
    \caption{Softmax simplices of out-of-distribution classes within a QCD/W/Top classifier. ($\{p_k\}_{k=1}^3$ denotes the corresponding softmax probabilities in classes of QCD, W, and Top.)\label{fig:sm_simplex}}
\end{figure}

\begin{figure}
     \centering
     \includegraphics[width=0.32\textwidth]{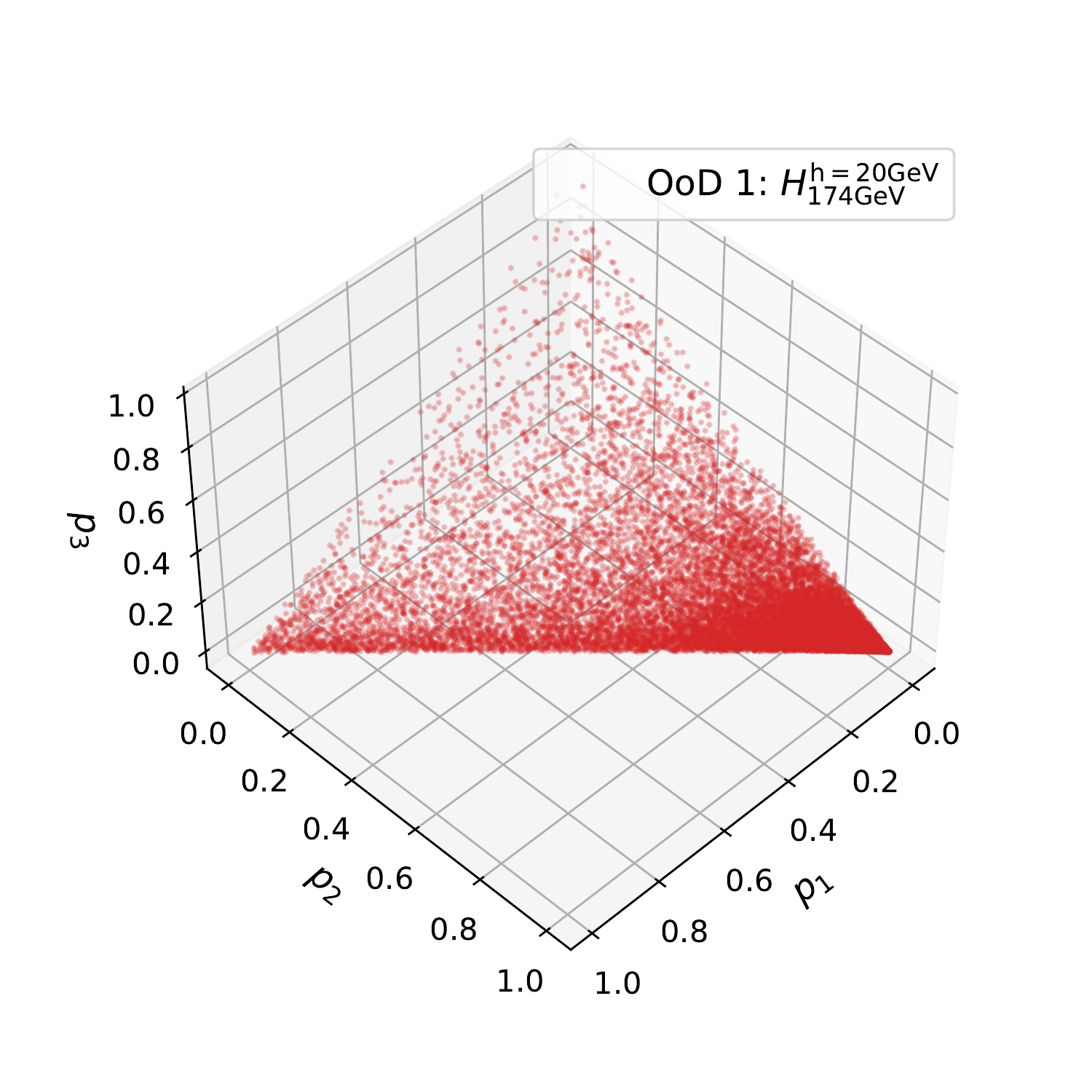}
     \includegraphics[width=0.32\textwidth]{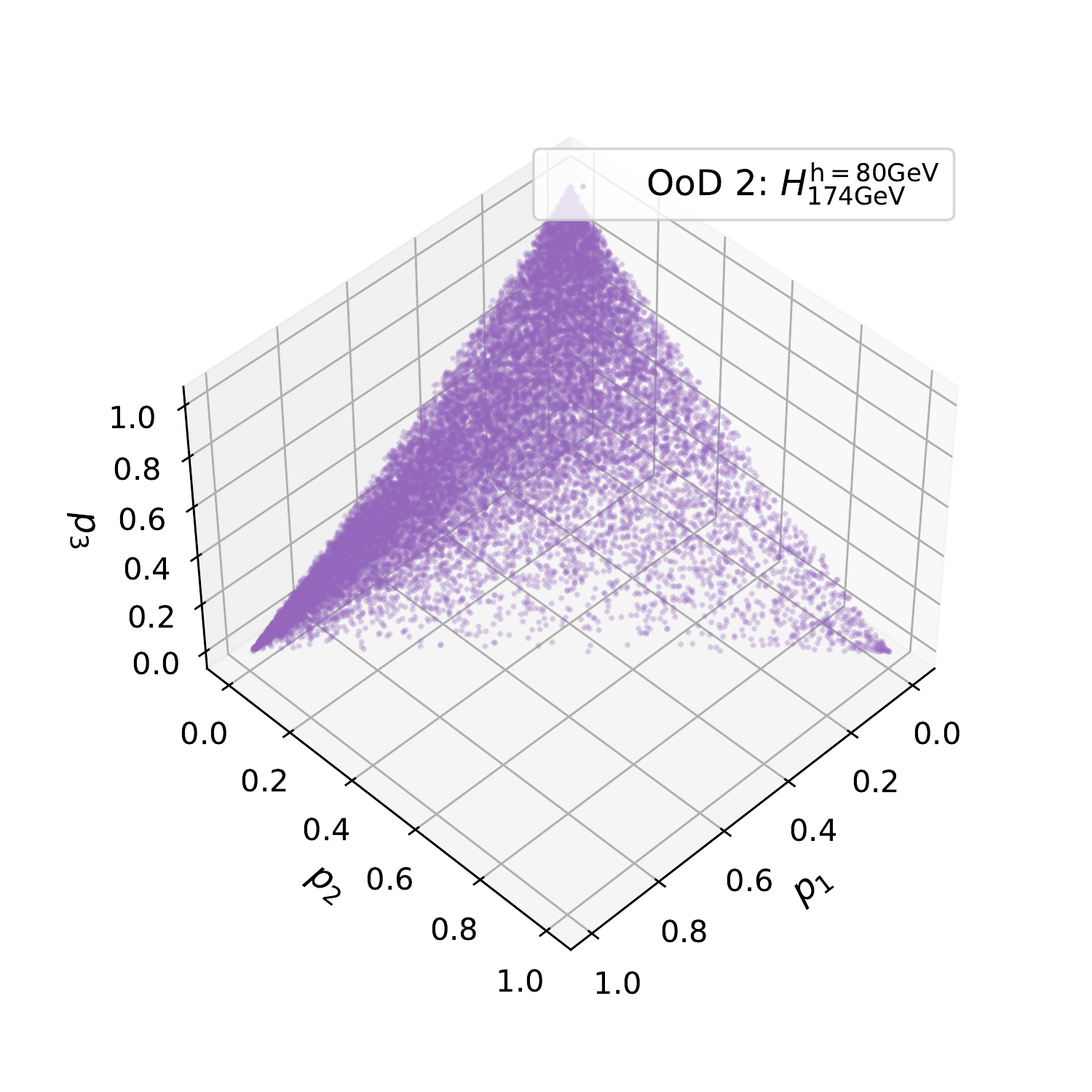}
     \includegraphics[width=0.32\textwidth]{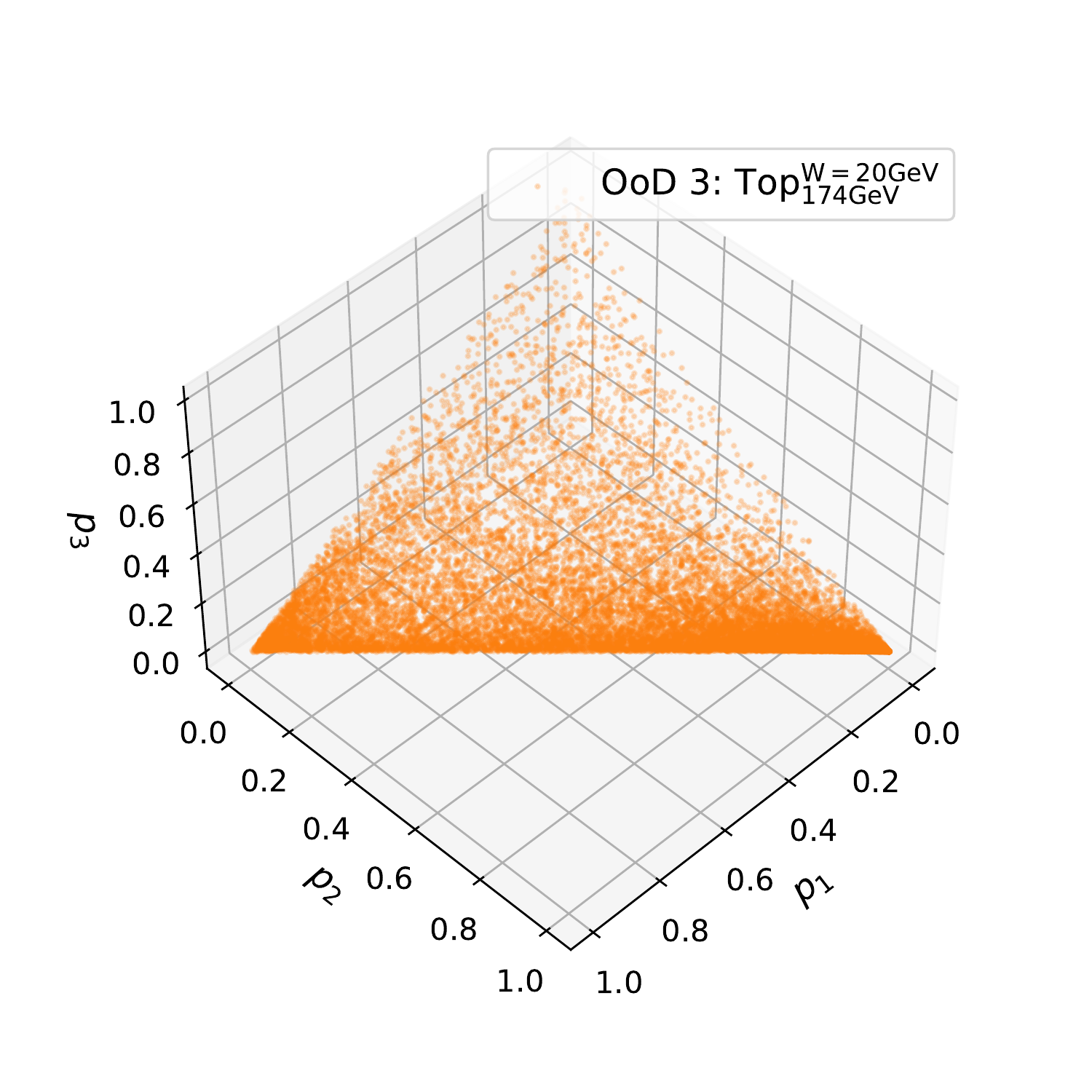}
    \caption{Softmax simplices of out-of-distribution classes within a mass-decorrelated QCD/W/Top classifier. ($\{p_k\}_{k=1}^3$ denotes the corresponding softmax probabilities in classes of QCD, W, and Top.)\label{fig:mdeco_sm_simplex}}
\end{figure}

Figure~\ref{fig:sm_simplex} shows the softmax simplex distributions for three different OoD signals. For in-distribution samples, they are expected to aggregate around the corners, which correspond to high classification confidence. While, for out-of-distribution samples, they are more randomly distributed on the simplex plane. More interestingly, the mass-decorrelation induces new patterns in the softmax simplices (figure~\ref{fig:mdeco_sm_simplex}). Generally speaking, without mass-decorrelation, the OoD examples are more QCD- and Top-like. With mass-decorrelation, the distributions change drastically, indicating different representation patterns. We will see later in section~\ref{sec:results} how this results in effective signal detection.

In practice, we can either cast the anomalous jet tagging problem as a standard OoD detection problem (i.e., inclusively against all the SM jets), or reformulate it as a class-conditional anti-QCD tagging problem. In the context of new physics searches, we have highly-imbalanced experimental data due to the hadronic nature of the LHC\@. Most of the background events are from QCD jet production processes. And other background processes can be effectively reduced or estimated. Thus in the following, we will focus the study on the formulation of anti-QCD tagging (as illustrated in the experiment in section \ref{subsec:exp}).

In principle, the anomaly scoring function can be composed with the predictive softmax probabilities ($p(\rvx)$ in eq.~(\ref{eq:sm_prob})), the logits ($\rvg (\rvx)$) before the softmax normalization, or the (penultimate) latent representations ($\rvh (\rvx)$).
\begin{equation}
\rvx \mapsto \rvh(\rvx); \, \rvg(\rvx) = \rmW \rvh(\rvx) + \rvb; \,
p_k (\rvx) = \frac{\exp(\rvg_k (\rvx))}{\sum_{k'}^K \exp(\rvg_{k'} (\rvx))}
\label{eq:sm_prob}
\end{equation}
For a well-calibrated model (more information in appendix~\ref{app:extra}), the classifier prediction can be safely interpreted as the posterior probability. Thus we can directly employ the predictive softmax probabilities as the anomaly scores. However, as was argued in~\cite{Hein2019WhyRN, Lee2018ASU}, the softmax outputs could be ``label-overfitted'' and overconfident for OoD data far away from the decision boundary. This observation motivates the usage of the penultimate layer representations, with which we can instead measure density probabilities in the feature space and be free of the overconfidence problem caused by piecewise affine transformations in Rectified Linear Unit (ReLU) -based networks. In the context of anti-QCD tagging, we compose (QCD) class-conditional scoring functions as follows:
\begin{itemize}
    \item {\bf Softmax probability based scoring function:} Class-conditional \emph{non-QCD} softmax probability $- p_{\rm QCD} (\rvx)$ with $p_{\rm QCD} = p(y=0 \mid \rvx)$ (the minus sign accounts for the reverse in \textbf{anti}-QCD detection).
    \item {\bf Representation based scoring function:} To directly measure the distance in latent space, we employ the Mahalanobis Distance~(MD)~\cite{Lee2018ASU, DBLP-ml-journals-ml-corr-ml-abs-2106-09022} in the penultimate layer, by viewing the classifier as a Gaussian model. Normally we calculate the mean and covariance matrix (eq. (\ref{eq:md2})) in the training set $\{\rvx_i^{\rm train}\}_{i=1}^N$, and measure the distance (eq. (\ref{eq:md})) from the test datum $\rvx$ to the target cluster. For anti-QCD tagging, $\{\rvx_i^{\rm train}\}_{i=1}^N$ refers to the QCD jets exclusively.
    \begin{subequations}
        \begin{equation}
            \texttt{MD} = (\rvh(\rvx) - \mathbf{\mu})^\top \Sigma^{-1}(\rvh(\rvx) - \mathbf{\mu})\,,
            \label{eq:md}
        \end{equation}
        \begin{equation}
            \textrm{with} ~ \mu = \frac{1}{N} \sum_i^N \rvh(\rvx_i^{\rm train}), ~ \Sigma = \frac{1}{N}\sum_i^N (\rvh(\rvx_i^{\rm train}) - \mu)(\rvh(\rvx_i^{\rm train}) - \mu)^\top
             \label{eq:md2}
        \end{equation}
    \end{subequations}
\end{itemize}

\subsection{Improving model uncertainty estimates}

The property of having higher uncertainty on unseen out-of-distribution examples is expected from a deep model to make reliable predictions in real-world settings. As improving uncertainty estimates~\cite{bib10.1007-ml-3-540-45014-9_1, Minderer2021RevisitingTC, pmlr-v70-guo17a, bib10.1145-ml-1102351.1102430} at the same time helps with OoD detection, we explore three different methods in this section.

\subsubsection{Deep ensemble}
Deep ensemble~\cite{bib2016arXiv161201474L} as a practical method for uncertainty estimation has been leveraged for OoD detection as well. Compared with Monte-Carlo Dropout~\cite{gal2016dropout}, it provides a more convenient protocol for uncertainty estimation. For classification problems, as indicated in eq.~(\ref{eq:ensemble}), an ensemble model is well-approximated with averaging the predictive probabilities of M individually trained classifiers $p^{(m)}(y \vert \rvx; \theta_m)$ with model weights $\theta_m$.
\begin{equation}
    p(y \vert \rvx; D) = \mathbb{E}_{p(\theta \vert D)} [p(y \vert \rvx; \theta)] \simeq \frac{1}{M} \sum_{m=1}^{M} p^{(m)}(y \vert \rvx; \theta_m)
    \label{eq:ensemble}
\end{equation}

\subsubsection{OVA-AVA combinational classification}

In the training of a classifier-based anomaly detector, the all-vs-all~(AVA) classification might have the downside that the decision boundary is not informative enough, i.e., the decision boundary is not fully aligned with the data manifold boundaries. In the setting of a closed softmax probability simplex, OoD examples, as well as the InD samples, are restricted to the plane. When OoD detection is fully driven by the decision boundary, it may lead to misspecification and high detection error. One way to amend this is to combine one-vs-all~(OVA) binary classification with all-vs-all classification~\cite{DBLP-ml-journals-ml-corr-ml-abs-2006-00954, DBLP-ml-journals-ml-corr-ml-abs-2007-05134}. The multiplicative combination of the OVA and AVA probabilities (resulting in a \emph{pseudo-probability} in the sense that it's not a real probability in theory) is then used as the anomaly scoring function as written in eq.~(\ref{eq:ova-ava}), where $p^{k-{\rm OVA}}(\rvx)$ denotes the one-class probability of the $k$-th OVA neural classifier (classifying non-$k$/$k$ (with labels 0/1) classes with model weights $\theta_k$) and $p_k^{\rm AVA}(\rvx)$ denotes the softmax probability for the $k$-th in-distribution class from the AVA classifier (with model weights $\theta$) for datum $\rvx$.
\begin{subequations}
\begin{equation}
    p_k^{\rm OVA-AVA}(\rvx) = p^{k-{\rm OVA}}(\rvx) \cdot p_k^{\rm AVA}(\rvx)\,,
\end{equation}
\begin{equation}
    \textrm{with} ~
p^{k-{\rm OVA}}(\rvx)  = p(y=1 \vert \rvx; \theta_k),~p_k^{\rm AVA}(\rvx) = p(y=k \vert \rvx; \theta)
\end{equation}
\label{eq:ova-ava}\relax
\end{subequations}

\begin{figure}
\centering
     \centering
     \includegraphics[width=0.45\textwidth]{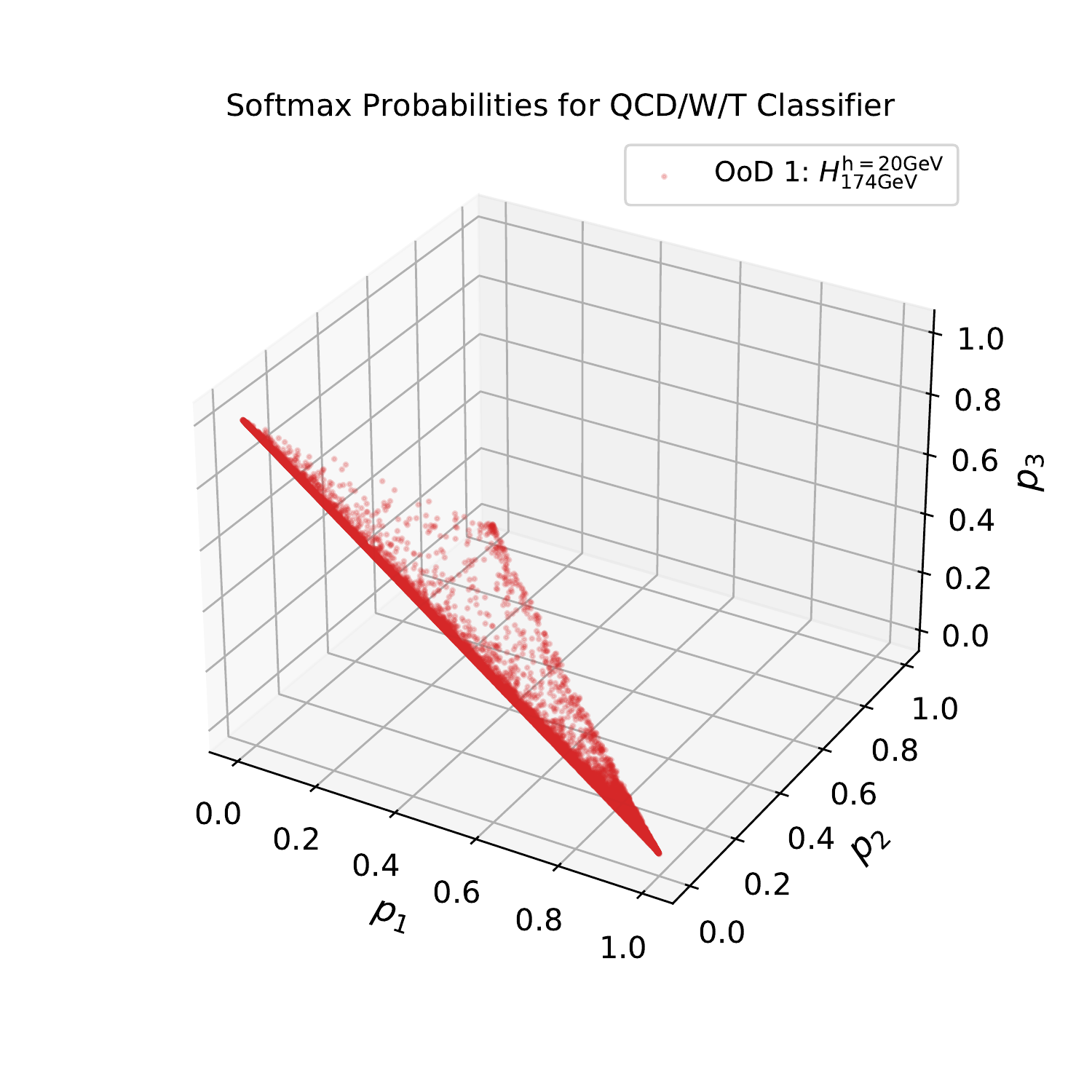}
     \includegraphics[width=0.45\textwidth]{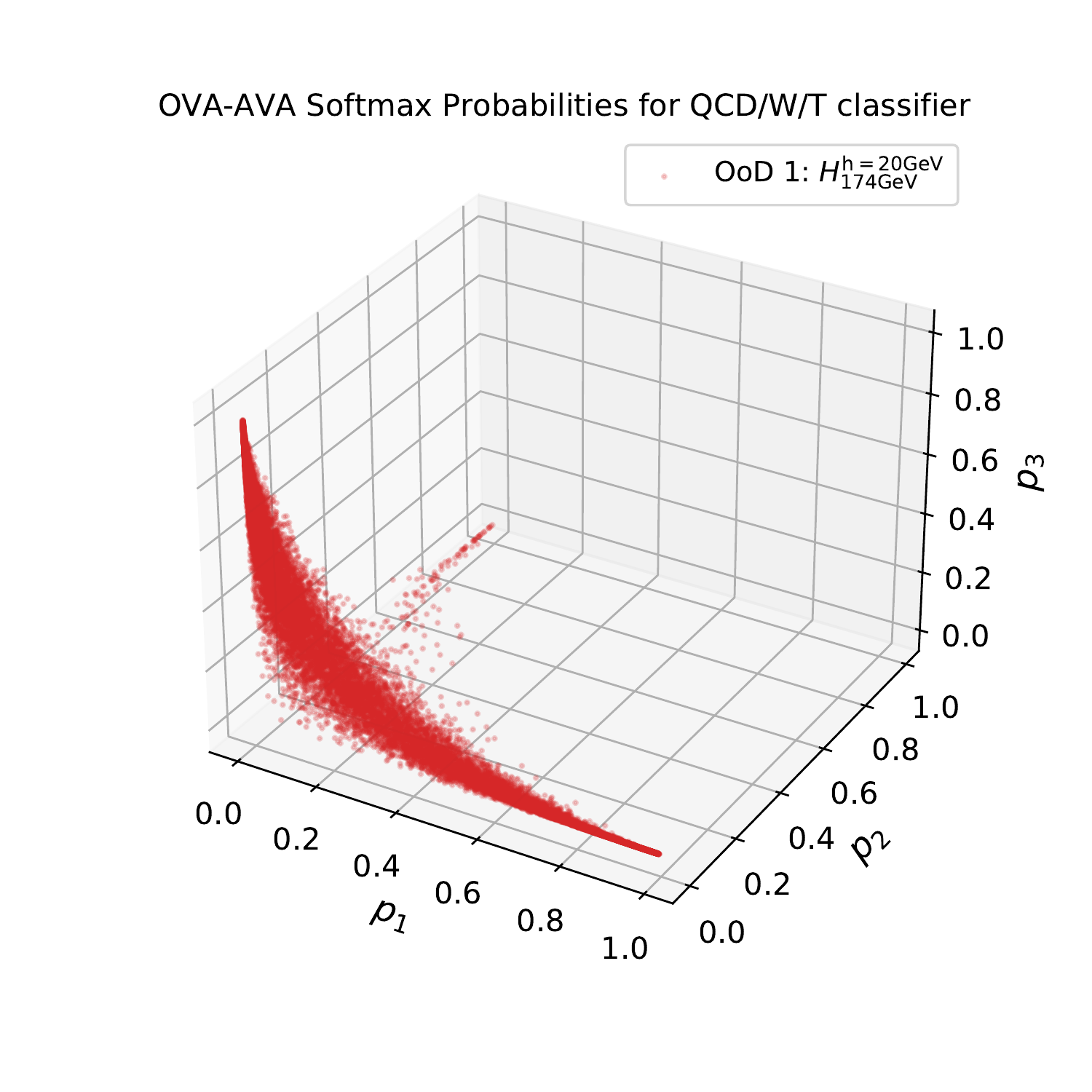}
    \caption{Softmax probabilities of out-of-distribution examples within the 
    standalone AVA classifier (\textit{left}) and the
    combined OVA-AVA scenario (\textit{right}).\label{fig:sm_ovaava}}
\end{figure}

In figure~\ref{fig:sm_ovaava}, we show the softmax probability simplex of one OoD class for a stand-alone all-vs-all classifier and the expanded simplex for the combined OVA-AVA scenario. The OVA factor brings InD and OoD examples even further apart. There are two forces within this action: 1) the all-vs-all classifier pulls OoD data points towards the center (corresponding to the uniform softmax distribution); 2) the one-vs-all classifiers pull OoD points away from the softmax plane towards the origin. At the same time, ideally for the in-distribution classes, $p^{\rm k-OVA}(\rvx_{\rm InD})$ and $p_k^{\rm AVA}(\rvx_{\rm InD})$ both approach 1 for the $k$-th corresponding class. This will result in better separation between the InD and OoD data.

\subsubsection{Distance-aware uncertainty estimate}

As mentioned previously, ReLU-based networks could be brittle and have high predictive confidence even for OoD examples far away from the InD samples.
It was proved that Radial Basis Function~(RBF) networks can mitigate this problem~\cite{Hein2019WhyRN}.  Gaussian Processes~(GP) with RBF kernels have the property of input distance awareness~\cite{bib10.5555-ml-1162254}, in the sense that RBF kernels tend to have uniform predictions ($p(y \vert \rvx) \sim \frac{1}{K}$) for far OoD examples.

In the same spirit, distance-aware uncertainty estimates can be achieved by a Spectral Normalized Gaussian Process~(SNGP)~\cite{DBLP-ml-journals-ml-corr-ml-abs-2006-10108} with two main ingredients: 1) the output layer replaced by a Gaussian Process with the RBF kernel:
\begin{equation}
    \rmK_{ij} = \exp(-\Vert \rvh(\rvx_i) - \rvh(\rvx_j) \Vert^2/2)\,,
    \label{eq:rbf}
\end{equation}
and 2) distance-preserving spectral normalization~\cite{DBLP-ml-journals-ml-corr-ml-abs-1802-05957} applied to the hidden layers.
Spectral normalization is used to control the Lipschitz constant of the classifier function by constraining the spectral norm layer by layer. Combining these two ingredients, we expect the distances between input instances to be well conserved while propagating through the neural net layers. The modified classifier can thus have larger uncertainty for OoD instances that are far away from the InD samples. In practice, the output Gaussian Process is implemented with a Laplace Approximation to the Random Fourier Feature~\cite{bib10.5555-ml-1162254} posterior of the GP.

\section{Setup}
\label{sec:setup}

\subsection{Datasets}
\paragraph{Training sets.}
The training set consists of boosted Standard Model QCD, $W$, and top jets with $\pt \in [550, 650]$\,GeV\@. The events are generated with \texttt{MadGraph}~\cite{Alwall_2011} for the 13\,TeV LHC, with the production processes $p p \to j j ~(j = u,d,c,s,g)$ for QCD jets, $p p \to W' \to W Z$ ($m_{W'}=1.25$\,TeV) for \W~jets, and $p p \to Z' \to t \bar t$ ($m_{Z'}=1.5$\,TeV) for top jets. The \W boson and top quark are restricted to hadronic decay modes ($Z$ bosons are restricted to decaying into neutrinos for collecting clean \W~jets). Generated events are then fed into \texttt{Pythia8}~\cite{Sj_strand_2008} and \texttt{Delphes}~\cite{de_Favereau_2014} for parton shower and fast detector simulation (with no pile-up effects simulated). We take particle flow objects for jet clustering, with no jet trimming applied. All jets are then clustered with the \texttt{anti-$k_{\rm T}$} algorithm~\cite{Cacciari_2008} with a cone size of $R=1.0$.  We have 350,000 jets for each class, with 20\% of the training samples serving as the validation~set.

\paragraph{OoD test sets.}
Test signal sets~\cite{cheng_taoli_2021_4614656} are hypothetical new physics jet types including boosted scalar jets (4-prong) and altered top jets (3-prong).  We borrow Two Higgs Doublet Models~(THDMs)~\cite{Branco-ml-2011iw} to generate boosted Higgs jets with pair production $p p \to H H$ ($H \to h(bb) h(bb)$), where the light Higgs boson is in the $h \to b \bar b$ decay mode. We set $m_H = 174$\,GeV with $m_h = 20, ~80$\,GeV to the Higgs bosons to express different shapes of ``four-prongness''.
In practice, we employ $h_3$ (requiring the parton-level $\pt > 600$\,GeV) in the THDM as the heavy Higgs, and $h_1$ as the light Higgs.
For the altered top jets, we rescale the intermediate \W~mass to 20\,GeV\@. A lighter \W~ will generate different relative radiation patterns for the altered top jets. The production process is the same as that for SM top jets.
Notation-wise, the OoD test classes are denoted by $H_{174 \GeV}^{h = 20 \GeV} $ ($m_H = 174$\,GeV with $m_h = 20$\,GeV), $H_{174 \GeV}^{h = 80 \GeV}$ ($m_H = 174$\,GeV with $m_h = 80$\,GeV) and $\textrm{Top}_{174 \GeV}^{W = 20 \GeV}$ ($m_{\rm Top} = 174$\,GeV with $m_W = 20$\,GeV).

All jets are again clustered using the \texttt{anti-$k_T$} algorithm with the cone size of $R=1.0$. Test jet $\pt$s are confined to a narrow region [550, 650] GeV for a fair comparison. Each test set with refined $\pt$ contains 20,000~samples.

As an extra note, we only test on non-SM OoD jets, since we focus on the out-of-distribution detection performance here. For Standard Model W and Top jets, the tagging is equivalently a procedure of usual classification, since they are in-distribution classes.

\paragraph{Preprocessing and input format.}
All jets are preprocessed following the procedure in ref.~\cite{Cheng-ml-2020dal}, including centering and rotating in the $(\eta, \phi)$ plane to align the jet principal axes. The first 100 jet constituents with the highest $\pt$s are selected as the neural network inputs with coordinates $\{(\log E_i, \log p_{{\rm T}i}, \eta_i, \phi_i)\}_{i=1}^{100}$.

\subsection{Neural classifier}

ParticleNet~\cite{Qu-ml-2019gqs} is employed as the classifier architecture in this study. ParticleNet is a dynamic convolutional neural net constructed on the $k$ nearest neighbours of each jet constituent. It performs excellently in multiple jet classification tasks including QCD/Top, QCD/W, and quark/gluon classification. The categorical cross-entropy ($-\sum_k^K y_k \log(p_k(\rvx))$ where $y_k$ denotes the one-hot encoding of the label) is used as the loss function for the QCD/W/Top 3-class classification. 

\subsection{Training settings}

We adopt the same hyper-parameters and learning rate scheduling from~\cite{Qu-ml-2019gqs} for the standard training and the OVA-AVA scenario. The model is trained for 30 epochs with the Adam optimizer~\cite{kingma2017adam}. Convergence is encouraged by a decreasing learning rate which follows a 1-cycle learning rate schedule \cite{Smith2019SuperconvergenceVF}. (Check appendix~\ref{app:arch} for more details.)

\paragraph{Deep ensemble.}
For classification tasks, deep ensembling practically amounts to training M models independently and averaging over the predictive probabilities. We train 10 independent models with random initial weights and data orderings for the ensemble model.

\paragraph{OVA-AVA.}
To perform anti-QCD tagging with the combined classification, we train a QCD/W/Top AVA 3-class classifier and a QCD/(W, Top) OVA binary classifier with the same training samples.

\paragraph{SNGP.}
For the SNGP training, we re-optimize the learning rate and schedule based on the procedure in~\cite{Smith2018ADA}.  More experimental details can be found in appendix~\ref{app:arch}.

\section{Results}
\label{sec:results}

We present the performance of classifier-based anomalous jet tagging in this section. We first examine the un-decorrelated tagger in generic jet tagging. Then we present the mass correlation and decorrelation effects and their impacts on OoD detection.

\subsection{Anomalous jet tagging}
At inference time, the trained classifiers are used to calculate anomaly scores to perform binary classification on in-distribution QCD jets and OoD signal jets. By thresholding the post hoc anomaly scores, we select jets with the highest scores and identify them as non-QCD signals.
To facilitate model comparison and estimate the overall performance, we employ the Receiver Operating Characteristic~(ROC) Curve and the Area Under the ROC Curve (AUC) as metrics.

As introduced in section~\ref{sec:methods}, we investigate different training strategies and model setups. We explore 1) the Single Model, 2) the Deep Ensemble with 10 independent runs, 3) the OVA-AVA combinational model, and 4) the Distance-aware SNGP model. The softmax probability-based scoring function is simply calculated as $- p_{\rm QCD} (\rvx) = - p(y=0 | \rvx)$ while QCD jets are labelled in 0. The $p_{\rm QCD} (\rvx)$ is averaged over all the 10 runs for the emsemble model. For the OVA-AVA model, $p_{\rm QCD}^{\rm OVA-AVA}$ is calculated as $p^{\rm OVA}_{\rm QCD}(\rvx) \cdot p_{\rm QCD}^{\rm AVA}(\rvx)$.
As for the class-conditional Mahalanobis distance, we extract the penultimate layer representations $\rvh(\rvx)$ and calculate the class mean and covariance matrix (eq.~(\ref{eq:md2})) with 100,000 in-distribution QCD training samples. Though the Deep Ensemble and the OVA-AVA model are motivated in the space of softmax probabilities, we can still calculate the corresponding MDs in practice.
For Deep Ensembles, the MD is averaged over the individual models. And for the OVA-AVA model, we sum up the MDs from the OVA classifier and the AVA classifier.

\begin{table}
    \centering
    \resizebox{\textwidth}{!}{
    \begin{tabular}{ccccc}
    \hline
     \backslashbox{Scenario}{Signal}   & &   $H_{174 \GeV}^{h = 20 \GeV} $ & $H_{174 \GeV}^{h = 80 \GeV}$ & $\textrm{Top}_{174 \GeV}^{W = 20 \GeV}$\\ \hline
        \multirow{2}{*}{SingleModel}
                                & $p_{\rm QCD}$
                                & 0.782 $\pm$ 0.004 & 0.872 $\pm$ 0.001 & 0.768 $\pm$ 0.004 \\
                                & \MD
                                & 0.840 $\pm$ 0.008 & 0.856 $\pm$ 0.006 & 0.815 $\pm$ 0.008 \\
    \hline
    \multirow{2}{*}{Ensemble10}
                                & $p_{\rm QCD}$  & 0.786 $\pm$ 0.001  & \textbf{0.878} & 0.774 $\pm$ 0.001 \\
                                & \MD & 0.848 $\pm$ 0.001 & 0.866 $\pm$ 0.001 & \textbf{0.824 $\pm$ 0.001} \\
    \hline
    \multirow{2}{*}{OVA-AVA}
                                & $p_{\rm QCD}$     & 0.784 $\pm$ 0.002 & 0.876 $\pm$ 0.001 & 0.772 $\pm$ 0.002 \\
                                & \MD & \textbf{0.852 $\pm$ 0.005} & 0.861 $\pm$ 0.003 & 0.822 $\pm$ 0.004 \\
    \hline
    \multirow{2}{*}{SNGP}
                                & $p_{\rm QCD}$                                                                 & 0.786 $\pm$ 0.004 & 0.875 $\pm$ 0.001 & 0.777 $\pm$ 0.003 \\
                                & \MD
                                & 0.834 $\pm$ 0.005 & 0.860 $\pm$ 0.006 & 0.819 $\pm$ 0.005 \\
    \hline
    \end{tabular}}
    \caption{OoD test AUCs for different model training strategies and anomaly scoring functions ($p_{\rm QCD}$ and Mahalanobis distance \MD). Standard deviations are calculated from 10 random runs, except that for the Ensemble model we bootstrap for 5 times from 20 individual models. (In the case of a too-small ($<0.001$) standard deviation, we omit it in the table.) For each test signal, the highest AUC is highlighted in~bold.\label{tab:aucs}}
\end{table}

\begin{figure}
    \centering
    \includegraphics[width=0.45\textwidth]{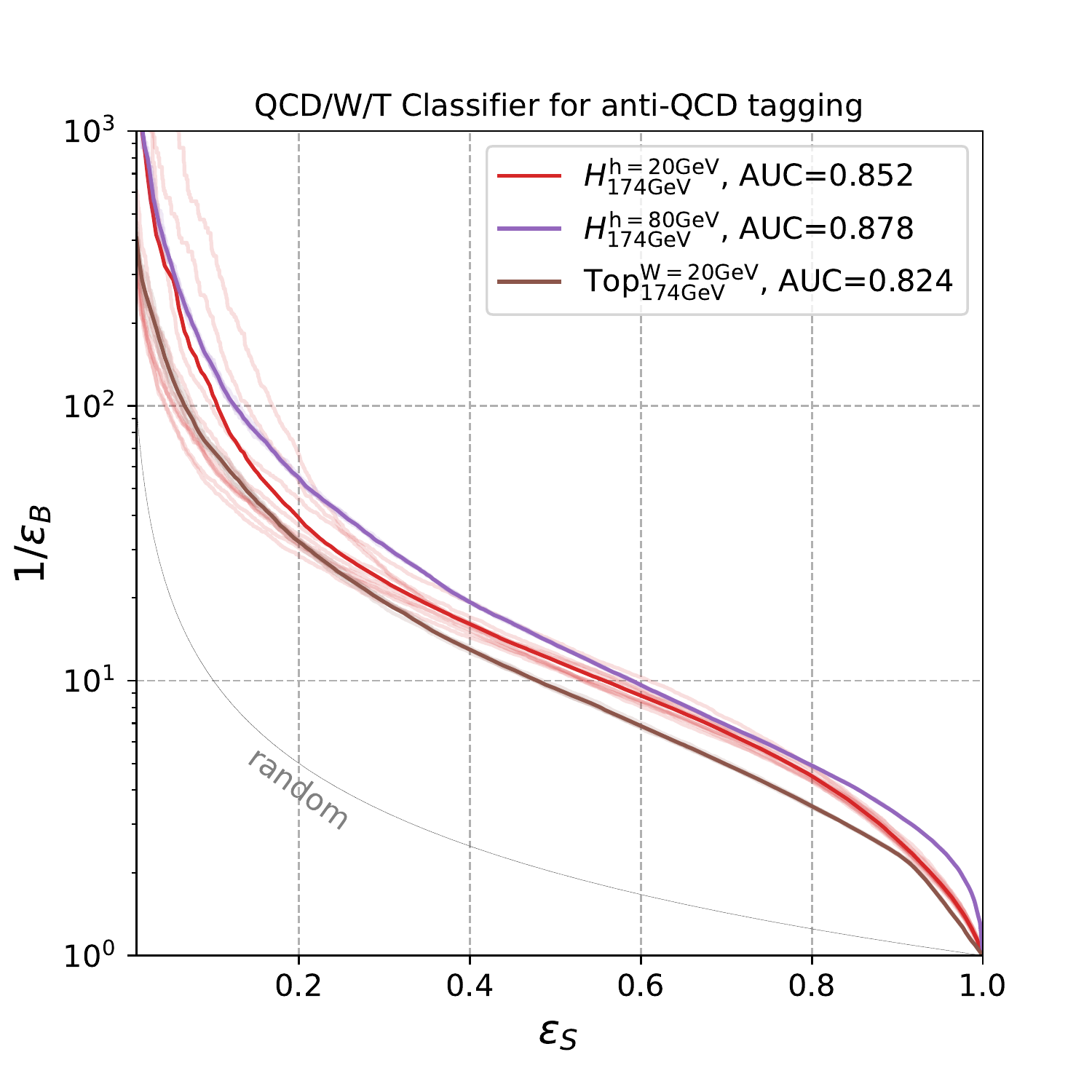}
    \includegraphics[width=0.45\textwidth]{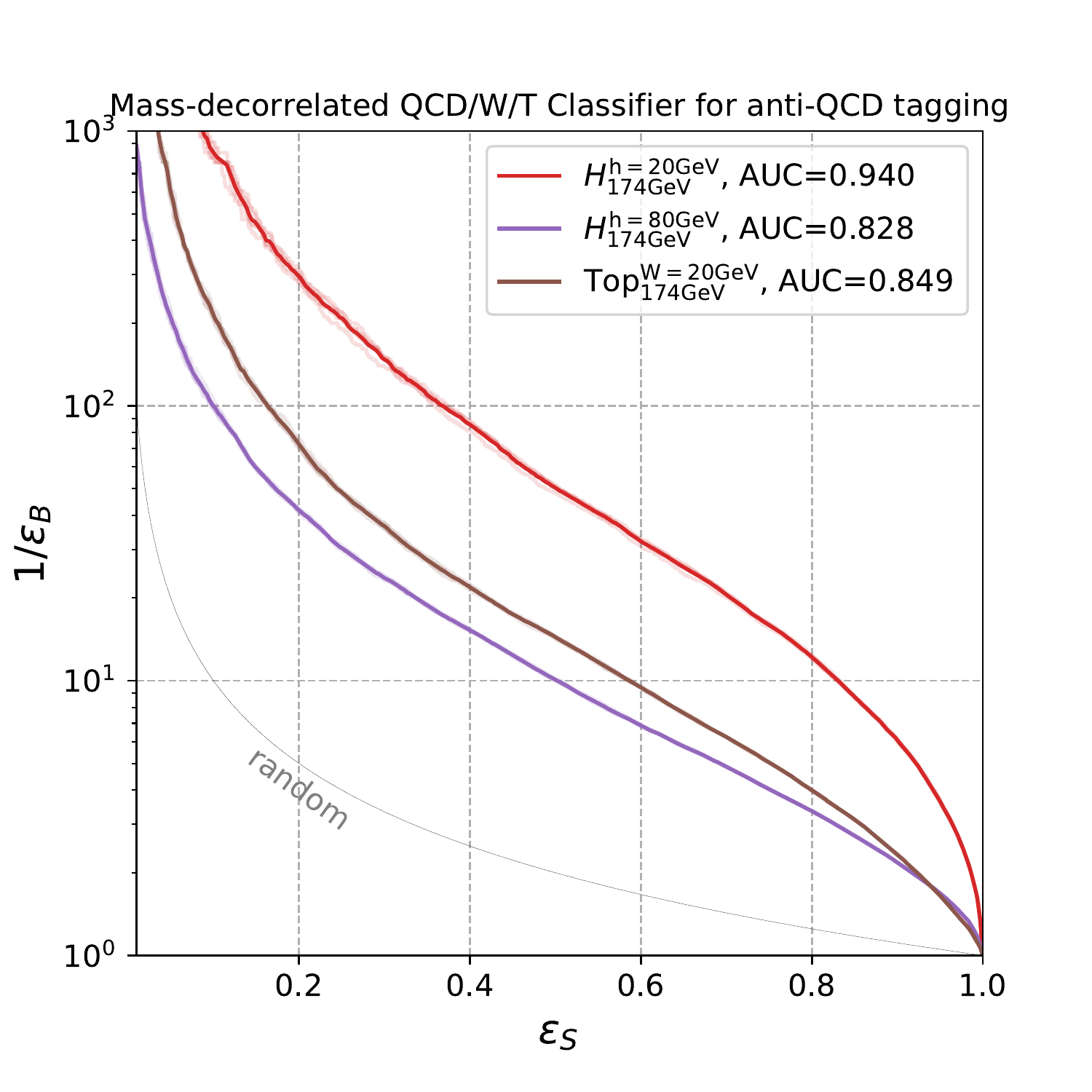}
    \caption{OoD ROC curves for the un-decorrelated QCD/W/Top classifier (\emph{left}) and the mass-decorrelated QCD/W/Top classifier (\emph{right}).\label{fig:rocs}}
\end{figure}

AUCs for discriminating test signals against the QCD background are recorded in table~\ref{tab:aucs}. Standard deviations are calculated from 10 independent runs with random initial weights and data orderings. Corresponding ROC curves for the optimal scenarios are shown in figure~\ref{fig:rocs}. In summary, we have the following observations:
\begin{itemize}
    \item We reach AUCs $\sim 0.8-0.9$ for all the test signals, corresponding to background rejection rates $1/\epsilon_B (\epsilon_S=0.5) \sim 9.4-13.5$ and significance improvement factors $\epsilon_S/\sqrt{\epsilon_B} \sim 1.53-1.84$.
    \item Different scenarios have different sensitivity regions. The latent-representation-based scoring function outperforms the softmax-probability-based scoring function in 2 out of 3 test signals, while $p_{\rm QCD}$ performs slightly better for $H_{174 \GeV}^{h = 80 \GeV}$.
    Relatively speaking, $H_{174{\rm GeV}}^{\rm h = 20\,GeV}$ (with a very light $m_h$, it has two highly boosted sub-jets, which will further decay, and thus could be similar to a W jet in the high-level topology) and ${\rm Top}_{174 \GeV}^{W = 20 \GeV}$ are closer to the in-distribution classes. Thus we can approximately categorize them as Near-OoDs~\cite{Winkens2020ContrastiveTF}.
            \item The three improvement methods for uncertainty estimates all slightly strengthen the tagging performance. It confirms that better uncertainty estimation helps with OoD detection.
\end{itemize}

Though not recorded here, we also observe that a better feature extractor (a neural architecture with a higher classification accuracy) comes along with better OoD detection performance (A Fully Connected Network~(FCN) is employed for model comparison. Results can be found in table~\ref{tab:fcn} of appendix~\ref{app:extra}).

\subsection{Mass decorrelated anomalous jet tagger}

To better serve general resonance searches and facilitate model comparison, we need to decorrelate the jet mass from the anomaly scoring function. More concretely, a generic anomalous jet tagger will confront the problem of mass sculpting (i.e., the tagger distorts the mass distribution of background jets). It has been considered a common problem of generic anomalous jet taggers. However, in this section, we argue that mass-decorrelation in the classifier-based approach augments OoD tagging efficiency, due to the inductive biases within the neural architecture and the classification task.

A mass-decorrelated jet tagger, generally, will have decreased tagging efficiency, since it blocks out the discriminative information of jet mass.
For example, naive autoencoder-based anomalous jet taggers are strongly mass-correlated, resulting in over-simplified representations. Mass-decorrelated autoencoders, if not trained with augmenting strategies, in most cases have difficulties in preserving effective OoD detection performance.
However, for neural classifier-based anomalous jet tagging, the manifestation of the mass correlation and the underlying mechanism are completely different: 1) Masses of the in-distribution classes will directly determine the mass correlation pattern.
This might result in biased signal detection since jets with masses close to the mass peaks of the in-distribution classes are given higher probabilities. 2) The predominant mass correlation will obscure effective representation learning for other relevant discriminative features such as jet substructure. 3) Mass-decorrelation in the classifier-based anomaly tagger plays a different role for anti-QCD tagging, compared to a supervised jet tagger or a generative anti-QCD tagger. It separates the mass-dependent part and the mass-independent part of the classification,  and activates more generic features that are helpful in anomalous jet tagging.

To decorrelate the jet mass from the classifier-function-based anomaly scores, we employ the resampling strategy in~\cite{Cheng-ml-2020dal}. By mass-augmenting and resampling the Standard-Model-like W and Top jets and matching their mass distributions to that of the QCD training samples, the classifier should be blind to the discriminative information from the jet mass.  Mass-rescaled Stand Model W jets (250,000) and Top jets\footnote{For smaller Top masses ($M_t \leq ~80 \GeV$), the intermediate W boson mass is relatively~rescaled.} (160,000) are generated to span the full QCD mass spectrum. The neural architecture and the training procedure are the same as in the non-decorrelated version.

\begin{figure}
    \centering
    \includegraphics[width=\textwidth]{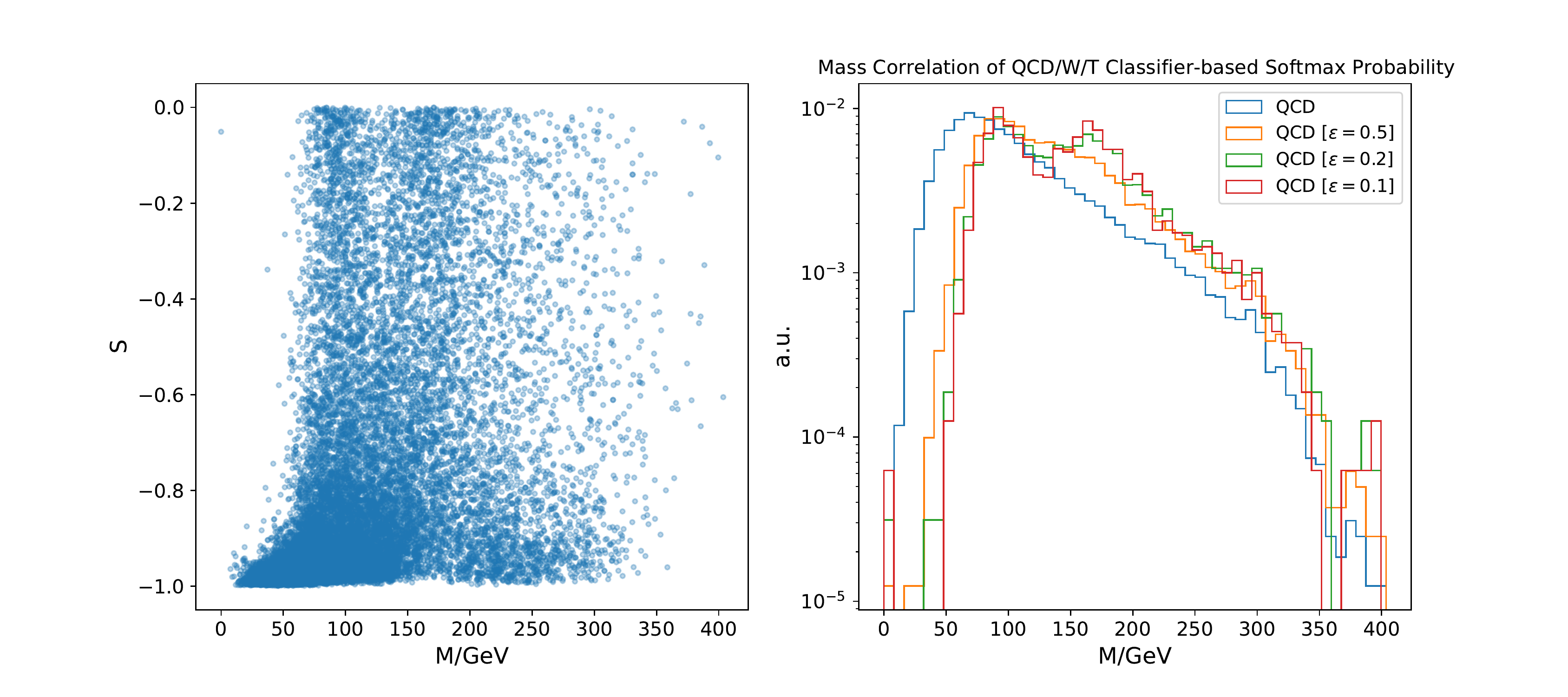}
    \includegraphics[width=\textwidth]{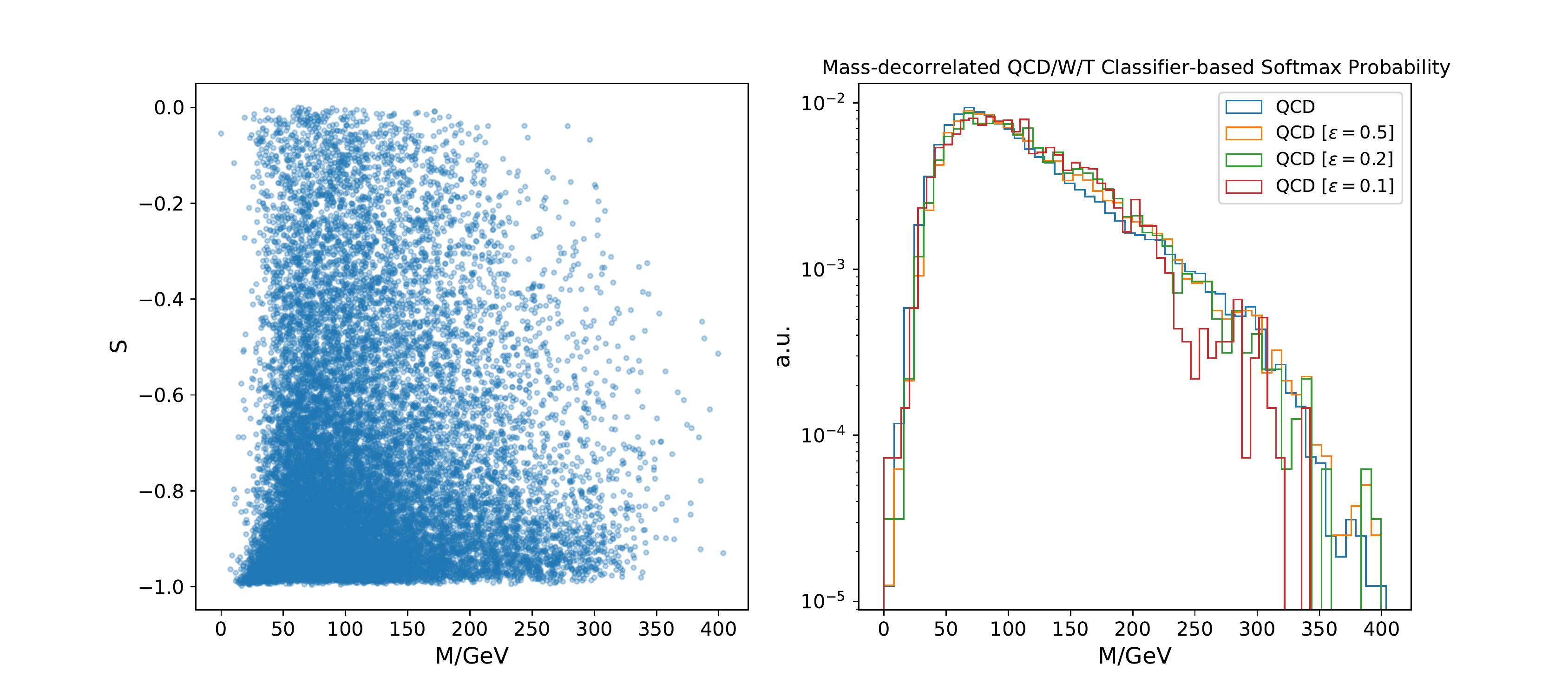}
    \caption{\emph{Top}: The un-decorrelated model with $p_{\rm QCD}$ as the anomaly score.
            \emph{Bottom}: The mass-decorrelated model with $p_{\rm QCD}$ as the anomaly score.
            \emph{Left}: The correlation between jet mass M and the anomaly score.
            \emph{Right}: Mass distributions for different background acceptance rates~($\epsilon$).\label{fig:mdeco_pqcd}}
\end{figure}

\begin{figure}
    \centering
    \includegraphics[width=\textwidth]{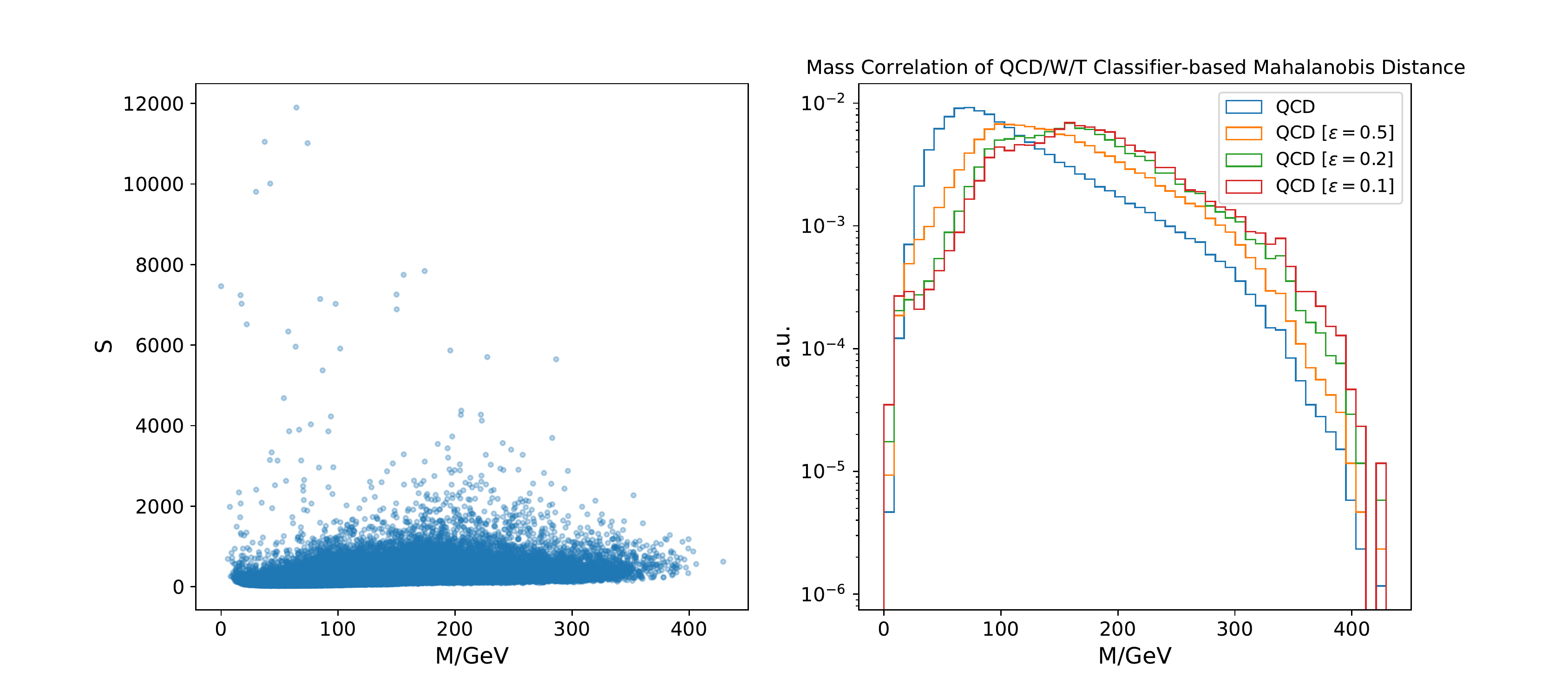}
    \includegraphics[width=\textwidth]{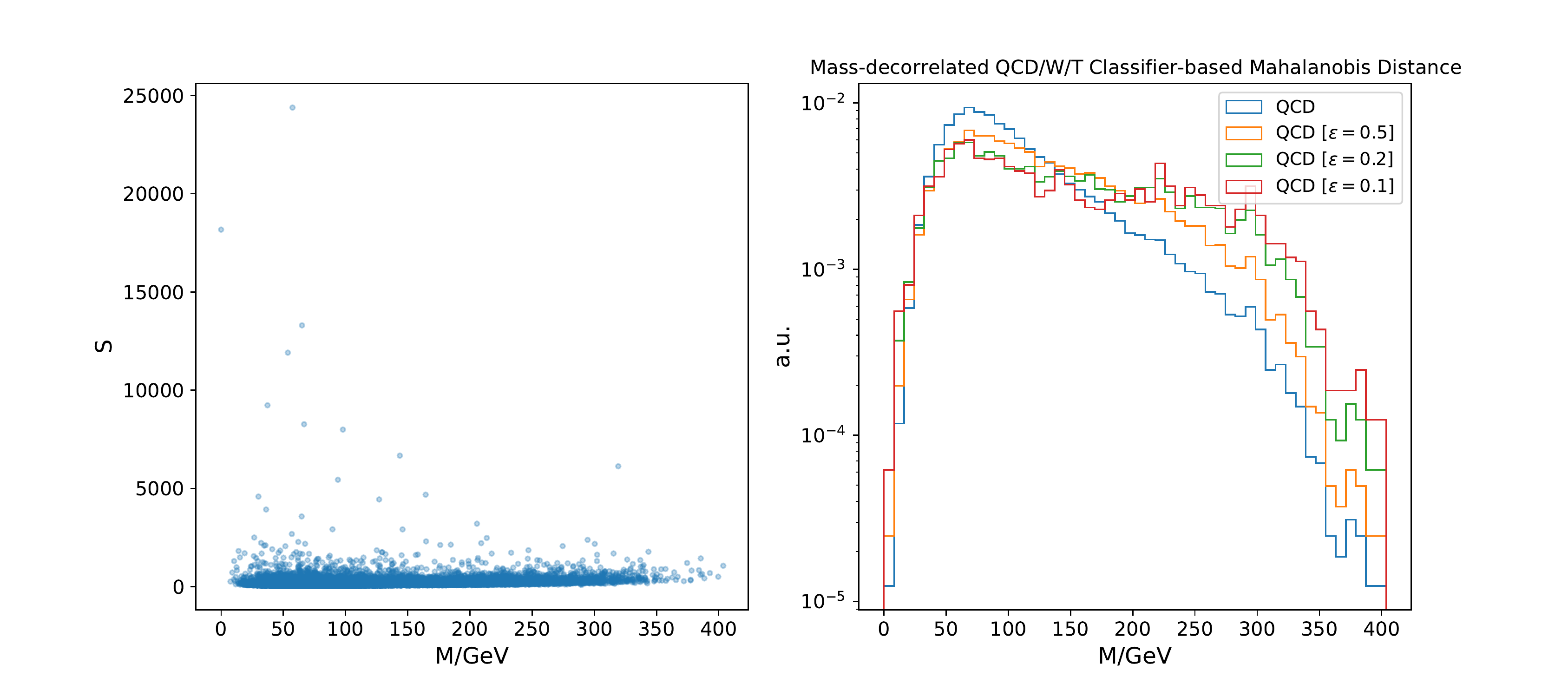}
    \caption{\emph{Top}: The un-decorrelated model with the Mahalanobis distance as the anomaly score.
            \emph{Bottom}: The mass-decorrelated model with the Mahalanobis distance as the anomaly score.
            \emph{Left}: The correlation between jet mass M and the anomaly score.
            \emph{Right}: Mass distributions for different background acceptance rates~($\epsilon$).\label{fig:mdeco_md}}
\end{figure}

\looseness=-1
Mass correlation and decorrelation results for the Single Model are shown in figure~\ref{fig:mdeco_pqcd} and~\ref{fig:mdeco_md} for $p_{\rm QCD}$-based and Mahalanobis distance-based scoring functions (more results can be found in appendix~\ref{app:extra}).  $p_{\rm QCD}$ is strongly shaped by in-distribution jet masses (peaking at the W and Top masses) and is perfectly mass-decorrelated when the classifier is trained with the mass-augmented dataset. However, the latent Mahalanobis distance operates under a slightly different mechanism.
We note that the non-decorrelated latent distance (\emph{Top Panel} in figure~\ref{fig:mdeco_md}) is weakly mass-correlated compared with $p_{\rm QCD}$, and is not strictly subject to data-imposed mass invariance. The softmax probability-based scores are better regulated in this respect, since they are directly manifested in the training objective and trained with Maximum Likelihood Estimation. In comparison, the latent representations are more brittle and more reluctant regarding incorporating invariance within. This is however understandable because there are extra degrees of freedom\footnote{Translation in logits $g_k(\rvx) \to g_k(\rvx) + a$ will not change the softmax probabilities $p(\rvx)$ (eq.~(\ref{eq:sm_prob})).} eaten in the softmax function.

\begin{table}
    \centering
    \resizebox{\textwidth}{!}{
    \begin{tabular}{ccccc}
    \hline
     \backslashbox{Scenario}{Signal}    &    &  $H_{174 \GeV}^{h = 20 \GeV} $ & $H_{174 \GeV}^{h = 80 \GeV}$ & $\textrm{Top}_{174 \GeV}^{W = 20 \GeV}$\\ \hline
    \rowcolor{Gray}
    OE-VAE    &  &  0.744 & 0.625  & 0.721 \\ \hline
    \rowcolor{Gray}
         & $p_{\rm QCD}$  & 0.595 & 0.740 & 0.648 \\
    \rowcolor{Gray}
    \multirow{-2}{*}{DisCo($\lambda = 100$)} & \MD   & 0.645 & 0.745 & 0.681 \\
    \hline
    \multirow{2}{*}{SingleModel} & $p_{\rm QCD}$ & 0.937 $\pm$ 0.001 & 0.824 $\pm$ 0.001 & 0.846 $\pm$ 0.002 \\
         & \MD
         & 0.931 $\pm$ 0.005 & 0.757 $\pm$ 0.012 & 0.836 $\pm$ 0.008 \\
    \hline
    \multirow{2}{*}{Ensemble10}  & $p_{\rm QCD}$ & \textbf{0.940}  & \textbf{0.828}   & \textbf{0.849}  \\
                & \MD & 0.941 $\pm$ 0.001 & 0.770 $\pm$ 0.004 & 0.851 $\pm$ 0.001 \\
    \hline
    \multirow{2}{*}{OVA-AVA} & $p_{\rm QCD}$ & 0.938 $\pm$ 0.001 & 0.825 $\pm$ 0.001 & 0.848 $\pm$ 0.001 \\
                            & \MD  & 0.933 $\pm$ 0.004 & 0.746 $\pm$ 0.013 & 0.831 $\pm$ 0.006 \\
    \hline
    \multirow{2}{*}{SNGP} &  $p_{\rm QCD}$   & 0.936 $\pm$ 0.001 & 0.820 $\pm$ 0.001 & 0.846 $\pm$ 0.001 \\
         & \MD & 0.920 $\pm$ 0.004 & 0.793 $\pm$ 0.006 & 0.833 $\pm$ 0.007 \\
    \hline
    \end{tabular}}
    \caption{OoD test AUCs under mass decorrelation for different model training strategies and anomaly scoring functions ($p_{\rm QCD}$ and Mahalanobis distance \MD). Standard deviations are calculated from 10 random runs, except that for the Ensemble model we bootstrap for 5 times from 20 individual models. (In the case of a too-small ($<0.001$) standard deviation, we omit it in the table.) For each test signal, the highest AUC is highlighted in bold.\protect\footnotemark Results from the mass-decorrelated Variational Autoencoders are shown as a reference. In addition, we also present a regularization-based mass decorrelation method (DisCo) as a~comparison.\label{tab:aucs_mdeco}}
\end{table}

\footnotetext{Since the  Mahalanobis distance is not reaching perfect mass  decorrelation, we focus on $p_{\rm QCD}$ here.}

The AUCs for mass-decorrelated models are recorded in table~\ref{tab:aucs_mdeco}, with the corresponding best-case ROC curves shown in the right panel of figure~\ref{fig:rocs}. We obtain promising signal tagging efficiencies for all the test sets (AUC $\gtrsim 0.8$ with the background rejection rates $1/\epsilon_{B}(\epsilon_S=0.5) \sim 10.2-51.0$). We also show the results from the Outlier Exposed Variational Autoencoders (OE-VAE)~\cite{Cheng-ml-2020dal} for model comparison. It's evident that the classifier-based approach outperforms OE-VAE.\footnote{We note that the encoding architecture in OE-VAE (mainly dense layers) is simpler than the architecture we used here. This leaves some room for further improvement in the OE-VAE~approach.}

Overall, $p_{\rm QCD}$ performs better than \MD, in both mass decorrelation and signal detection.
The Deep Ensemble model with the softmax probability as the anomaly score is the best scenario here.
Surprisingly, $H_{174{\rm GeV}}^{\rm h = 20\,GeV}$ even reaches an AUC of 0.940, which is exceptionally high for the current anomalous jet tagging benchmarks. At the signal acceptance rate of 50\%, the background rejection rate reaches 51.0 and results in a significance improvement factor of 3.6.
At the same time, mass decorrelation yields a different performance pattern in the test sets. Intriguingly, $H_{174{\rm GeV}}^{\rm h = 20\,GeV}$ and ${\rm Top}_{174{\rm GeV}}^{\rm W = 20\,GeV}$ have even higher AUCs than the corresponding un-decorrelated cases. This trend holds for both anomaly scores.
Normally for in-distribution classes, the mass-decorrelated classifier will have a slightly lower classification accuracy due to the deducted mass information.
Since the masses of the OoD samples are close to one of the in-distribution classes (Top of 174\,GeV), the different patterns between the InD and the OoD classes indicate that mass-decorrelation induces different  representations that are useful for OoD detection. (Recalling figure~\ref{fig:sm_simplex} and figure~\ref{fig:mdeco_sm_simplex}, the aggregation patterns in the un-decorrelated case are deeply affected by the jet mass, and the mass-decorrelated case is manifesting more substructures.)
The mass-decorrelated tagger benefits from the extra learning capacity while eliminating the predominant factor of jet mass.
We view $H_{174{\rm GeV}}^{\rm h = 20\,GeV}$ and ${\rm Top}_{174{\rm GeV}}^{\rm W = 20\,GeV}$ as Near-OoDs as introduced in the previous section.
Thus for Near-OoDs, mass decorrelation generally increases the AUCs, while Far-OoDs (e.g.\ typical 4-prong $H_{174{\rm GeV}}^{\rm h = 80\,GeV}$) have slightly lower AUCs compared to the non-decorrelated case.

In an extra investigation, we observe that the augmented performance could not be achieved with only decorrelating the jet mass. In table~\ref{tab:aucs_mdeco} we also present a regularization based mass decorrelation method: Distance Correlation (DisCo)~\cite{Kasieczka-ml-2020yyl} between the classifier prediction and the jet mass is added to the cross-entropy loss as a regularization term. We choose the regularization strength $\lambda = 100$ (see appendix~\ref{app:arch} for more details). The DisCo AUCs show that simply decorrelating mass doesn't result in improved OoD detection. The actual learning power is closely related to the data augmentation we employed here.

\paragraph{Data-augmented mass decorrelation improves OoD detection?}
It's intriguing that decorrelating the jet mass with data augmentation results in more performant OoD detection. The data augmentation process matches the mass distributions of the in-distribution classes (i.e., $p(M | y = k) = p(M) ~ \forall k$).
To better understand the underlying mechanism, we factorize the discriminative log-likelihood $\log p(y|\rvx)$ as
\begin{equation}
    \log p(y|\rvx) = \log p_{\cancel{M}}(y|\rvx) + \log p_M(y|\rvx)\,,
    \label{eq:mdeco_likelihood}
\end{equation}
where $\log p_M(y|\rvx)$ denotes the mass-dependent part and $\log p_{\cancel{M}}(y|\rvx)$ denotes the mass-independent part of the log-likelihood. This factorization is made possible by separating the mass from representation learning, since the mass distribution is invariant across the in-distribution classes.
Then we can view mass-decorrelation as a likelihood ratio approach~\cite{Ren2019LikelihoodRF}, with the mass-dependent part cancelled out.
At the same time, data augmentation in the mass dimension is inducing rich feature learning through the mass-independent part, and is expressing more subtle and weakly-correlated structures.
In general resonance searches (e.g., bump hunt), we can scan potential mass windows and thus bring back the mass dimension for identifying the mass bump.
 With the improved learning power, the overall performance is increased when we bring back the jet mass for discrimination. So we have not only reached mass decorrelation, but also further improved the OoD detection performance\footnote{The equivalent un-decorrelated tagging performance will be boosted compared with the classifier without this decorrelation~procedure.} in the procedure of decorrelating the jet mass. This phenomenon echoes with the improved ABCD method for better targeted signal detection~\cite{Kasieczka-ml-2020pil} and reveals the possibility of boosted model-independent search strategies.

\section{Discussion}
\label{sec:discussion}

\paragraph{Extension in the building blocks.}
We have established the general framework and demonstrated the potential of discriminative anti-QCD tagging. There are a few building blocks worth further exploring.
\begin{itemize}
    \item \textbf{Training datasets.} Currently we only train on the boosted Standard Model jets. It's possible that the tagging performance and coverage could be further improved by systematically including other potential hypothetical new physics signals.
    \item \textbf{Training strategy.} We employ the simplest training in the current setting, which is not specifically designed for OoD detection. We expect even improved performance with advanced training techniques. For instance, we can introduce regularization techniques and auxiliary tasks in the training process (e.g., the Outlier Exposure mentioned in the Introduction),
    or even pre-train the classifier on other larger datasets to further improve the representation learning capacity. Concurrently it's worth exploring training objectives other than the default categorical cross-entropy.
        \item \textbf{Domain adaptation.} The supervised classifier depends on simulation data at the moment. In order to apply to real-data analysis with less distortion, it's worth investigating the performance under distribution shifts and developing corresponding domain adaptation strategies~\cite{Ganin2016DomainAdversarialTO, Baalouch2019SimtoRealDA}. A common strategy is adversarial training distinguishing between simulation and real data.
     By making simulated and real QCD jets compete with each other, the goal is to learn domain-adapted representations and make the classifier prediction invariant over different data-generating settings.
\end{itemize}

\paragraph{Discriminative vs generative.}
Generative models have been employed as one important approach for anomalous jet tagging.
Although they could be a powerful tool for density estimation and accordingly be used for new signal detection, there are a few downsides associated. As one important motivation of our discriminative approach (or \emph{CLassiFier-based Anomaly Detection} --- CLFAD), it is revealed that the generative approach might assign higher likelihoods to OoD examples than to InD data. In contrast, the discriminative classifiers incorporate inductive biases that might help more effectively identify OoD instances.\footnote{However, we don't conclude that the classifier-based approach is completely guaranteed to be free from the misspecification~problem.}
We summarize the comparison between these two approaches in table~\ref{tab:comparison}.

\begin{table}
    \centering
    \begin{tabular}{m{6cm}|m{6cm}}
    \hline
    \textbf{Discriminative}  & \textbf{Generative} \\ \hline \hline
    Representation-driven     &  Likelihood-driven, density estimation\\
    \hline
    Extra freedom of in-distribution classes    &  Sensitive to dominant correlations (in the cases without further learning guidance) \\
    \hline
    Mass correlation depends on in-distribution classes & Strong mass correlation \\
    \hline
    Sensitive to jet types   & Possibility of assigning high likelihoods to OoD samples (observed in both computer vision and jet physics)\\
    \hline
    Simulation-Data domain adaptation  & Train directly on data\\
    \hline
    \end{tabular}
    \caption{\label{tab:comparison}Comparison between the discriminative and the generative approach in generic anomalous jet~tagging.}
\end{table}

Briefly speaking, the generative approach is based on the density estimation ability of generative models, while the discriminative approach leverages more the representation learning aspect of neural feature extractors. There are great efforts in the community to push the limit of powerful neural jet classifiers. Taking advantage of these sophisticated architectures, targeted jet classifiers and general model-independent new physics searches then support and motivate each other.
The selection of simulated in-distribution classes plays an important role in the discriminative approach, while we can train the generative models directly on data with negligible signal contamination (since they are supposed to be rare given the null results up to now). Both approaches demonstrate the mass correlation, however, with different underlying mechanisms.
As for the tagging performance and failure modes, CLFAD tends to have higher sensitivity to jet types, especially the similarity to in-distribution classes. Generative models risk missing signals in a large scope (\emph{perfect density estimation does not guarantee robust OoD detection}). Especially with mass-decorrelation, a (Variational) Autoencoder-based tagger could fail drastically for most test signals. Finally, as mentioned above, the simulated training set of CLFAD might have deteriorated performance under distribution shifts when deployed in real-data analyses. In this case, we need to additionally calibrate the neural models.

\section{Summary}
\label{sec:summary}

Despite the quick evolving of anomaly detection applications in LHC physics, the efforts are temporarily focused on a narrow track following generative models and density estimation. However, a discriminative neural classifier is by itself an anomaly detector in the sense that it encodes discriminative information equipped with inductive biases to tag anomalies. In this work, we have established the framework for discriminative anomalous jet tagging.

In building the framework, we leverage the SM particles as known in-distribution jet classes to perform multi-class classification. The trained classifier is then employed as an anomaly detector. A well-calibrated classifier will behave differently when confronted with OoD examples, either having higher predictive uncertainties or having latent representations far away from in-distribution clusters. Furthermore, viewing the classifier as a feature extractor, the latent representations should preserve information for discriminating between in-distribution and out-of-distribution samples.

In practice, we optimized an edge-convolution-based neural classifier for QCD/W/Top multi-class classification. At inference time, we investigated different anomaly scoring functions depending on the ingredients involved: 1) softmax probabilities and 2) penultimate latent vectors. The softmax probabilities are more stable and directly reflect the log-likelihood of the classifier.
The latent representations are trickier in the sense that they could contain more information than softmax probabilities, but they might be vulnerable and brittle.
Again, it has been discussed that model uncertainty estimation is linked with the performance of OoD detection. To better facilitate uncertainty estimation, we explored three approaches: 1) deep ensemble, 2) all-vs-all classification combined with one-vs-all classification, and 3) spectral normalized Gaussian Process.
To facilitate effective background estimation in general resonance searches, we augmented and resampled the training sets to match the mass distributions of all the in-distribution classes. In addition, mass decorrelation has another effect in this setup. By eliminating the predominant factor (i.e., the jet mass) in the learning process, the representation learning capacity is better exerted to manifest other discriminative factors.

For model evaluation, we employed hypothetical new physics particles (with significantly different radiation topologies) as test OoD signals. AUCs and ROCs are used for measuring the overall OoD detection performance. We have examined different scenarios, and from the experiments, we have observed that:
\begin{itemize}
    \item For all the test signals, we are able to reach AUCs larger than 0.8, no matter with mass decorrelation or not.
    \item The data-augmented and mass-decorrelated SM jet classifier can serve as a powerful generic anti-QCD jet tagger, thanks to the augmented feature learning. In the best case, we are able to reach an AUC of 0.940 and a background rejection rate of 51 at 50\% signal acceptance.
        \item The softmax probability-based scoring function is more stable compared with the latent representation-based scoring function. Due to the extra degrees of freedom, the latent Mahalanobis distance is less aware of data-imposed mass invariance. Overall, \emph{the mass decorrelated deep ensemble with $p_{\rm QCD}$ as the anomaly score is the scenario optimized for signal detection (with high sensitivity and precise mass decorrelation)}.
    \item A higher classification accuracy (thus a better feature extractor) and better uncertainty estimates help with OoD detection. Though in our case, the improvement effect from uncertainty estimation is not as significant.
\end{itemize}

In summary, we establish a framework for neural classifier-based anomaly detection in jet tagging for new physics searches at the LHC\@. Tailored for effectively reducing the most copious QCD background, we reframe the OoD detection problem as class-conditional anti-QCD tagging. We observe that with three ingredients: 1) a powerful feature extractor, 2) decent model calibration, and 3) data-imposed mass decorrelation, a classification-driven neural net can serve as a performant generic anti-QCD tagger. This observation paves for further studies on supervised classifier-based, however model-independent new physics~searches.

\acknowledgments

This work is supported by IVADO Postdoctoral Research Funding. Part of this work has been presented at the 2021 Machine Learning for Jets Workshop.

\appendix

\section{Neural nets and datasets}
\label{app:arch}

We employ the base model of ParticleNet~\cite{Qu-ml-2019gqs}. It consists of Edge Convolution blocks which convolve across 16 nearest neighbours in the $(\eta, \phi)$ plane. The Edge-Convolution features are aggregated by average pooling before being fed into the dense layers for classification. We follow the default settings (the architecture and the learning rate schedule) if not explicitly~stated.

\paragraph{ParticleNet}
\begin{description}
\item
  [\underline{\emph{Architecture}}]

  \texttt{Inputs}((2+2)$\times$100)$\to$ \texttt{EdgeConv}(64, 64, 64) $\to$ \texttt{EdgeConv}(128, 128, 128) $\to$ \texttt{EdgeConv}(256, 256, 256) $\to$ \texttt{Dense}(256) $\to$ \texttt{Softmax}(K). (K=3 for 3-class classification and K=2 for one-vs-all binary classification.)

\item
  [\underline{\emph{Hyper-parameters}}]
  The learning rate is scheduled as a 1-cycle procedure~\cite{Smith2019SuperconvergenceVF}: 3e-4 (8) 3e-3 (8) 3e-4 (4) 5e-7, where numbers in the brackets are interval epoch counts during which the learning rate is changed linearly. The batch size is set to 384. The models are trained for 30 epochs.
\end{description}

\paragraph{SNGP hyper-parameters.}
For the SNGP models, we 1) replace the output layer of ParticleNet with a Gaussian Process, and 2) apply Spectral Normalization to all the hidden layers.
We employ the hyper-parameters in table~\ref{tab:sngp_params}.

\begin{table}
    \centering
\renewcommand{\arraystretch}{1.2}
    \begin{tabular}{c|c}
\hline
Power Iteration & 1 \\
    Spectral Norm Bound (c) & 0.99\\ \hline
    Gaussian Process Hidden Dimension ($D_L$)     &  1024\\
    Length-scale Parameter ($l$)   & 1.0 \\
    $L_2$ Regularization  & 1e-6 \\
    Covariance Ridge Factor (r) & 1.00\\
    Covariance Discount Factor & 0.999 \\
    Learning Rate & 4e-5 (10) 4e-4 (10) 4e-5 (5) 1e-6 \\
\hline
\end{tabular}
    \caption{SNGP hyper-parameters. For the learning rate, we linearly increase it to the target value of 4e-4 and then decrease it to the initial value of 4e-5. Finally, we decrease the rate to 1e-6, and train for a few more epochs to ensure convergence. The numbers in brackets are epochs taken to finish the rate~transition.\label{tab:sngp_params}}
\end{table}

\paragraph{Distance correlation.}
The DisCo loss is written as follows:
\begin{equation}
    \Ls_{\rm DisCo} = \Ls (y, p(\rvx)) + \lambda~{\rm dCor}(M, p_{\hat k = k}(\rvx))\,,
\end{equation}
where $\Ls$ denotes the categorical cross-entropy loss and dCor is the distance correlation~\cite{bib2008arXiv0803.4101S}\footnote{We note that the definition dCor here is actually the squared distance correlation as defined in~\cite{bib2008arXiv0803.4101S}, for the sake of~simplicity.} between the jet mass M and the $\hat k$-th class prediction corresponding to the data class k. For the DisCo model, we train on the non-augmented data and fix the learning rate to 1e-3.

\section{Additional results}
\label{app:extra}

\paragraph{Model calibration.}

A well-calibrated classifier provides us with confident predictive probabilities for real-world deployment and downstream tasks. If the model prediction aligns with the true likelihood (i.e., the probability of an event being correctly classified ($\hat y = y$) is equal to the predicted confidence $\hat p$) as
\begin{equation}
    P(\hat y = y \mid \hat p = p) = p\,,
    \label{eq:calib}
\end{equation}
we then say the model is perfectly-calibrated.

\begin{figure}
    \centering
    \includegraphics[width=0.32\textwidth]{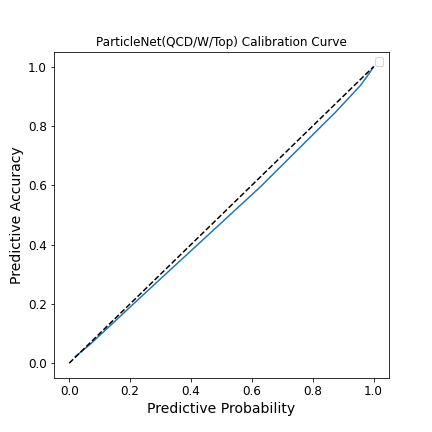}
    \includegraphics[width=0.32\textwidth]{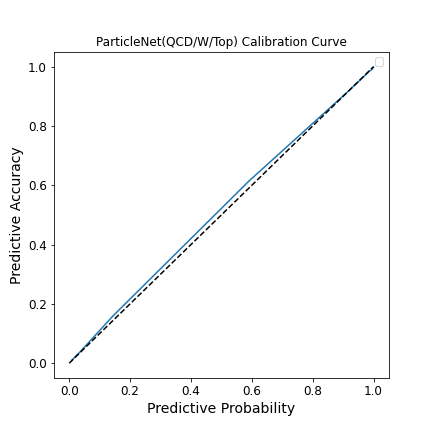}
    \includegraphics[width=0.32\textwidth]{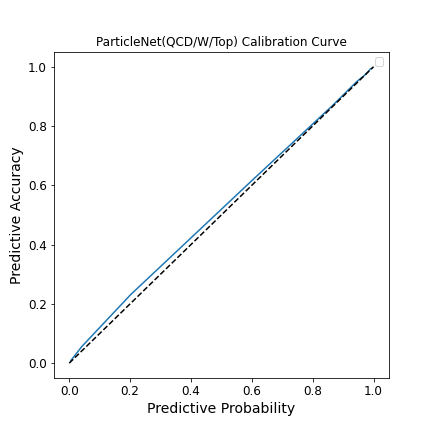}
    \caption{Calibration curves for ParticleNet trained on QCD/(W, Top) (\emph{left}), W/(QCD, Top) (\emph{middle}), and Top/(QCD, W) (\emph{right}).\label{fig:calib}}
\end{figure}

We have checked the calibration of ParticleNet.
We employ the Expected Calibration Error~(ECE)~\cite{Naeini2015ObtainingWC,pmlr-v70-guo17a} (defined as in eq.~(\ref{eq:ece})) as the metric,
where the classifier predictions are divided into M bins $\{B_m\}_{m=1}^M$, and \texttt{acc} denotes the accuracy and \texttt{conf} denotes the average confidence score in each bin. $|B_m|$ is the cardinality of the $m$-th bin, and n is the total sample size.
The ECE is calculated to be less than 1\% (0.6\% for 10 bins) for ParticleNet on the QCD/W/T dataset.
And we plot the calibration curve in figure~\ref{fig:calib} (the diagonal line corresponds to the perfect calibration case).
Thus under the current datasets and tasks, the model is well-calibrated.
\begin{equation}
    \textrm{ECE} = \sum_{m=1}^M \frac{|B_m|}{n} \lvert \textrm{acc}(B_m) - \textrm{conf}(B_m) \rvert
    \label{eq:ece}
\end{equation}

\paragraph{Fully connected network}
\begin{description}
\item
  [\underline{\emph{Input}.}] The four-vectors of $\pt$-ordered jet constituents $\{(E_i, p_{x,i}, p_{y,i}, p_{z, i})\}_{i=1}^{20}$ are taken as input features. They are properly normalized before being fed into the FCN.

\item
  [\underline{\emph{Architecture}.}] \texttt{Input}(4$\times$20) $\to$ \texttt{Dense}(256, ReLU) $\to$ \texttt{Dropout}(0.2) $\to$ \texttt{Dense}(128, ReLU) $\to$ \texttt{Dropout}(0.2) $\to$ \texttt{Dense}(12, ReLU) $\to$ \texttt{Softmax}(3)

\item
  [\underline{\emph{Hyper-parameters}.}]
\looseness=-1
The learning rate is scheduled as: 2.5e-4 (10) 2.5e-3 (10) 2.5e-4, where numbers in the brackets are interval epoch counts during which the learning rate is changed linearly. The batch size is set to 256. The model is trained for 50 epochs.
\end{description}

\begin{table}
    \centering
    \begin{tabular}{GccG}\hline
    Scenario    &  $H_{174 \GeV}^{h = 20 \GeV} $ & $H_{174 \GeV}^{h = 80 \GeV}$ & $\textrm{Top}_{174 \GeV}^{W = 20 \GeV}$\\ \hline
    FCN-$p_{\rm QCD}$     & 0.715 & 0.857 & 0.742 \\
    FCN-\texttt{MD}     & 0.747 & 0.730 & 0.773\\
    \hline
    \end{tabular}
    \caption{OoD test AUCs for a Fully Connected~Network.\label{tab:fcn}}
\end{table}

\paragraph{Extra mass (de)correlation plots.}

\begin{figure}[h]
    \centering
    \includegraphics[width=\textwidth]{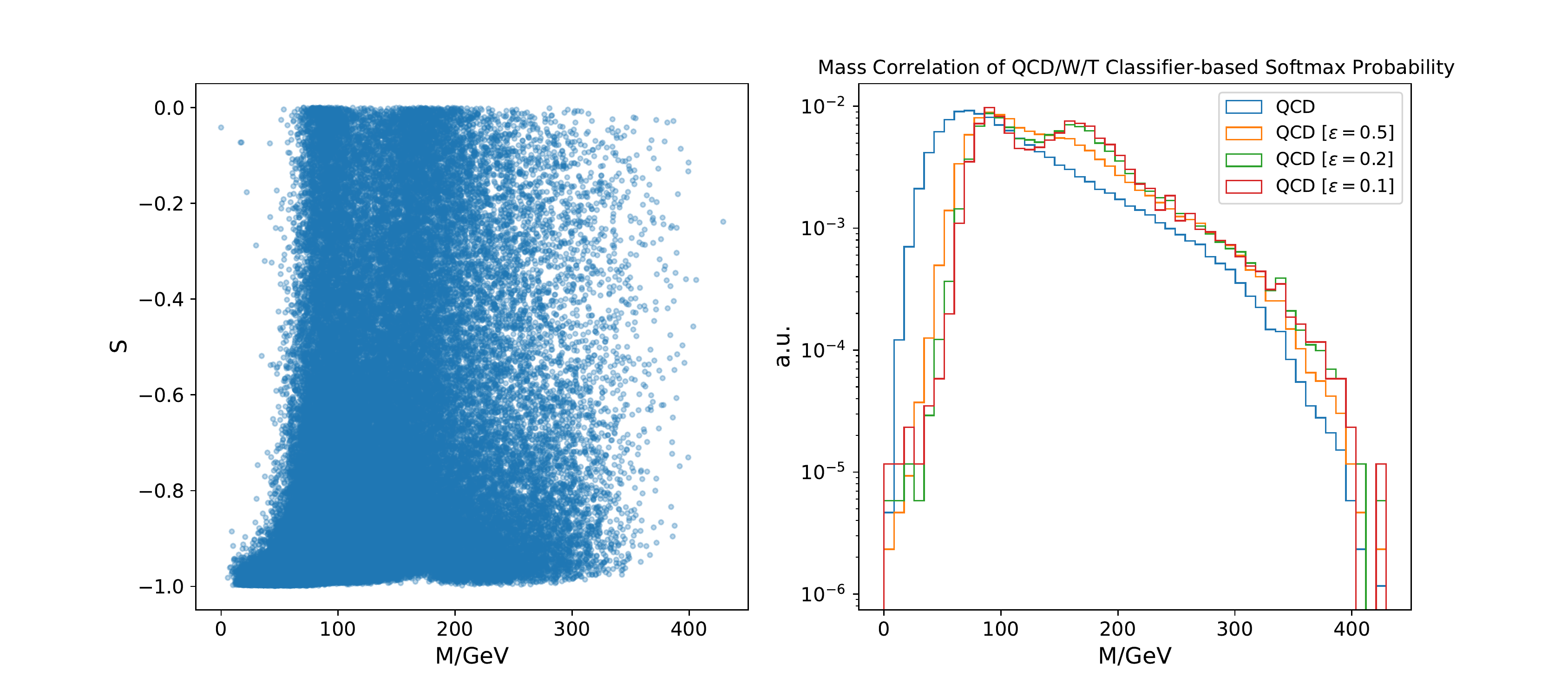}
    \includegraphics[width=\textwidth]{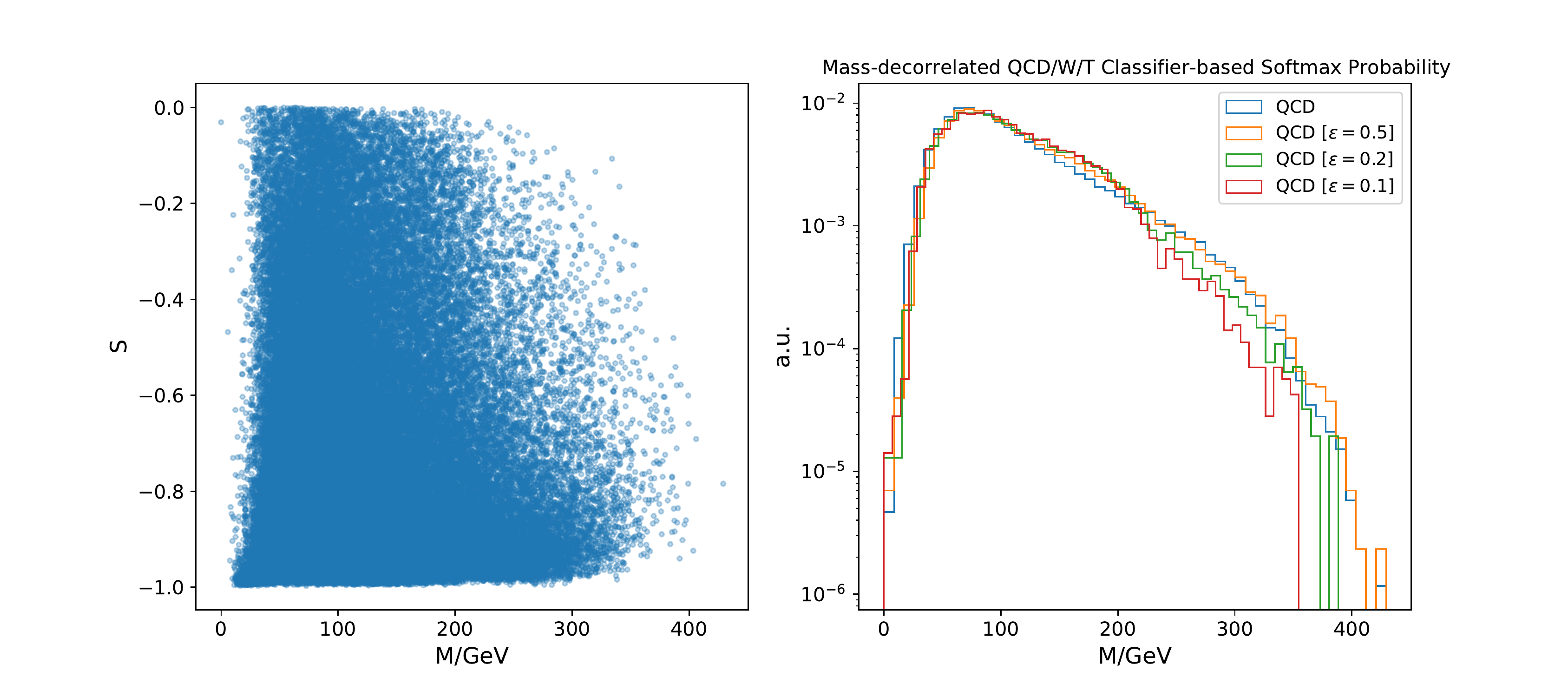}
        \caption{Ensemble Model.
            \emph{Top}: The un-decorrelated model with $p_{\rm QCD}$ as the anomaly score.
            \emph{Bottom}: The mass-decorrelated model with $p_{\rm QCD}$ as the anomaly score.
            \emph{Left}: The correlation between jet mass M and the anomaly score.
            \emph{Right}: Mass distributions for different background acceptance rates~($\epsilon$).\label{fig:mcorr_ensem}}
\end{figure}

\begin{figure}
    \centering
    \includegraphics[width=\textwidth]{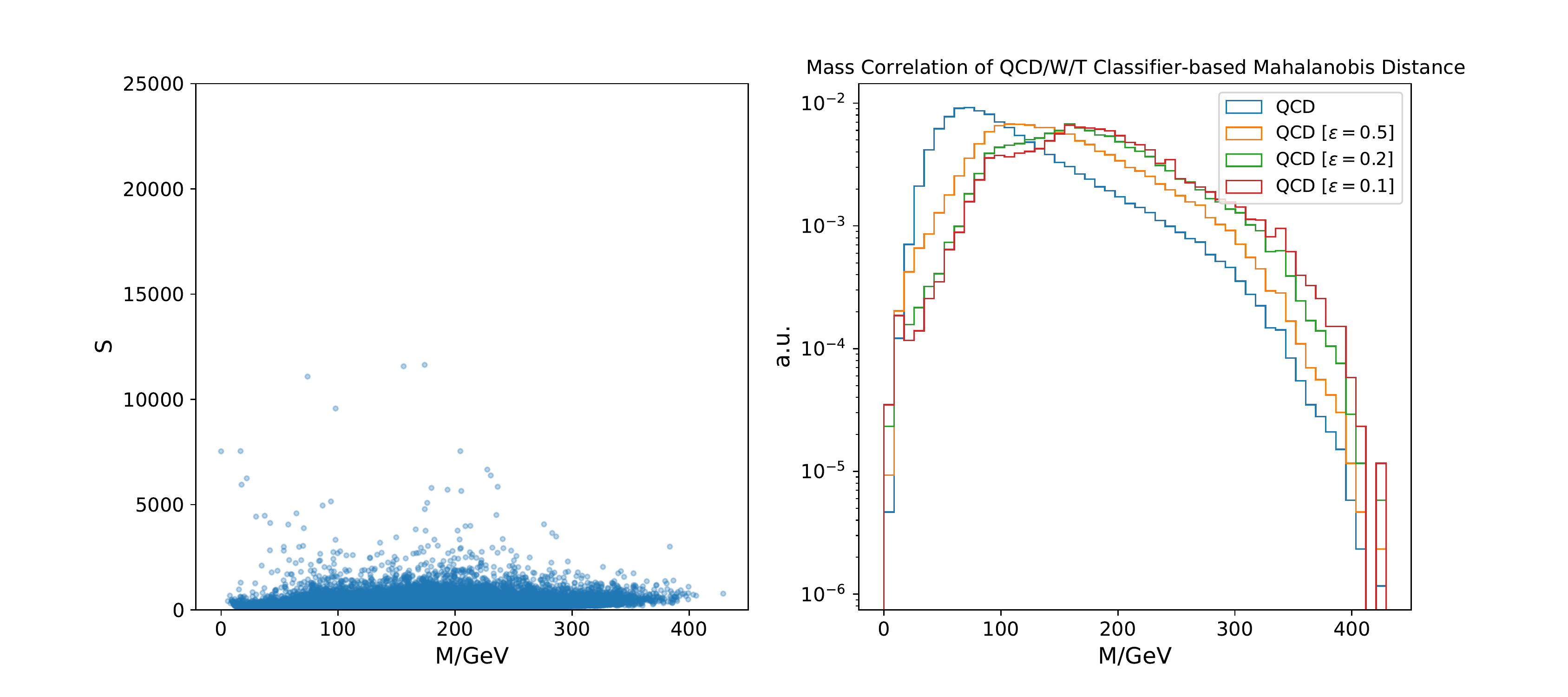}
    \includegraphics[width=\textwidth]{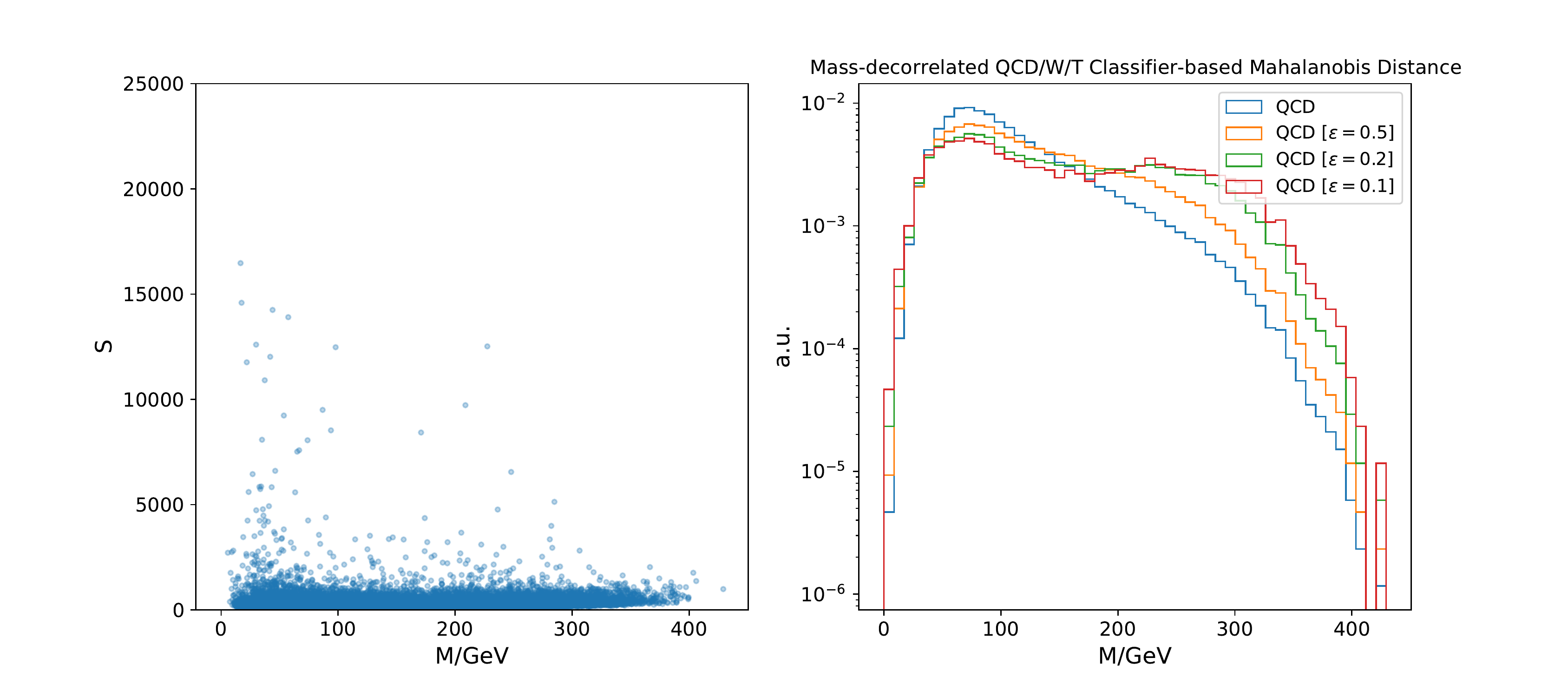}
        \caption{Ensemble Model.
            \emph{Top}: The un-decorrelated model with the Mahalanobis distance as the anomaly score.
            \emph{Bottom}: The mass-decorrelated model with the Mahalanobis distance as the anomaly score.
            \emph{Left}: The correlation between jet mass M and the anomaly score.
            \emph{Right}: Mass distributions for different background acceptance rates~($\epsilon$).\label{fig:mdeco_ensem}}
\end{figure}

\begin{figure}
    \centering
    \includegraphics[width=\textwidth]{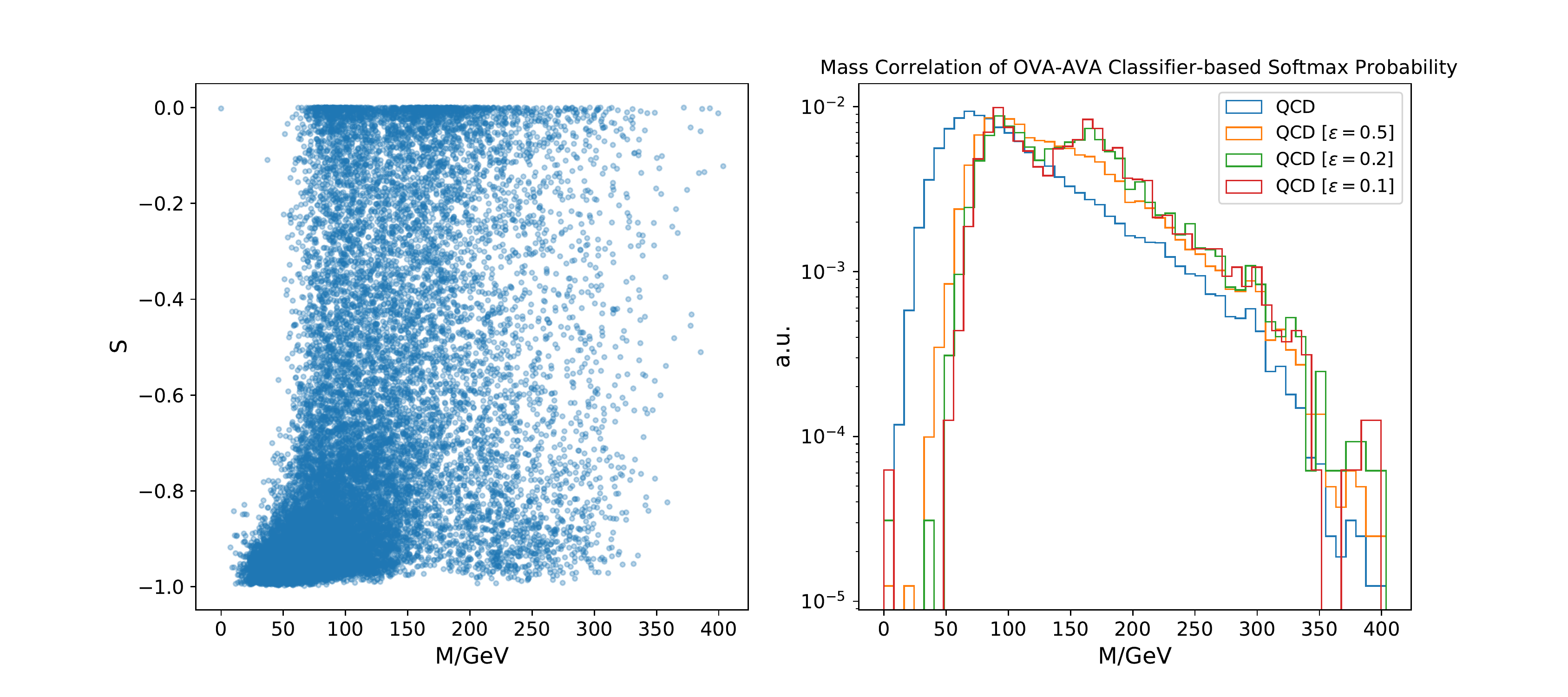}
    \includegraphics[width=\textwidth]{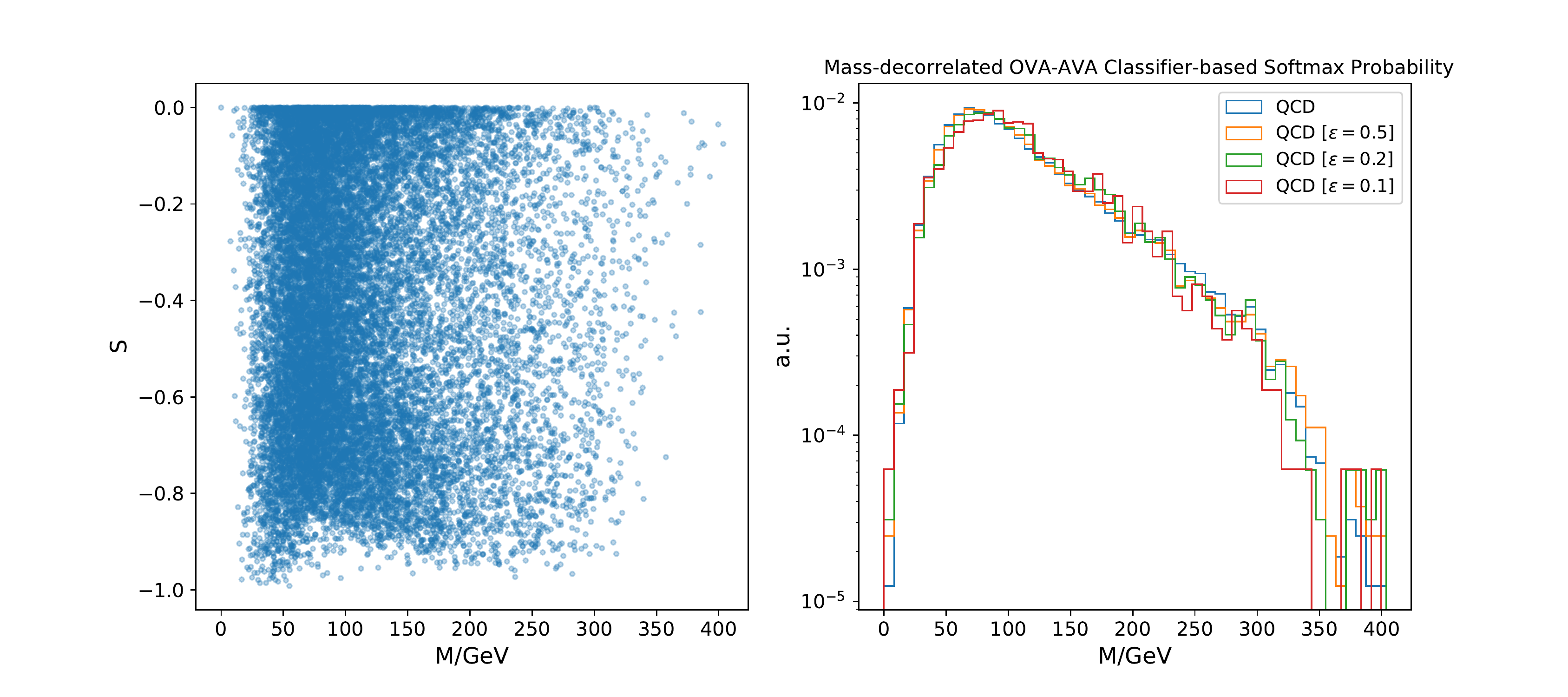}
        \caption{OVA-AVA Model.
            \emph{Top}: The un-decorrelated model with $p_{\rm QCD}$ as the anomaly score.
            \emph{Bottom}: The mass-decorrelated model with $p_{\rm QCD}$ as the anomaly score.
            \emph{Left}: The correlation between jet mass M and the anomaly score.
            \emph{Right}: Mass distributions for different background acceptance rates~($\epsilon$).\label{fig:mcorr_ovnni}}
\end{figure}

\begin{figure}
    \centering
    \includegraphics[width=\textwidth]{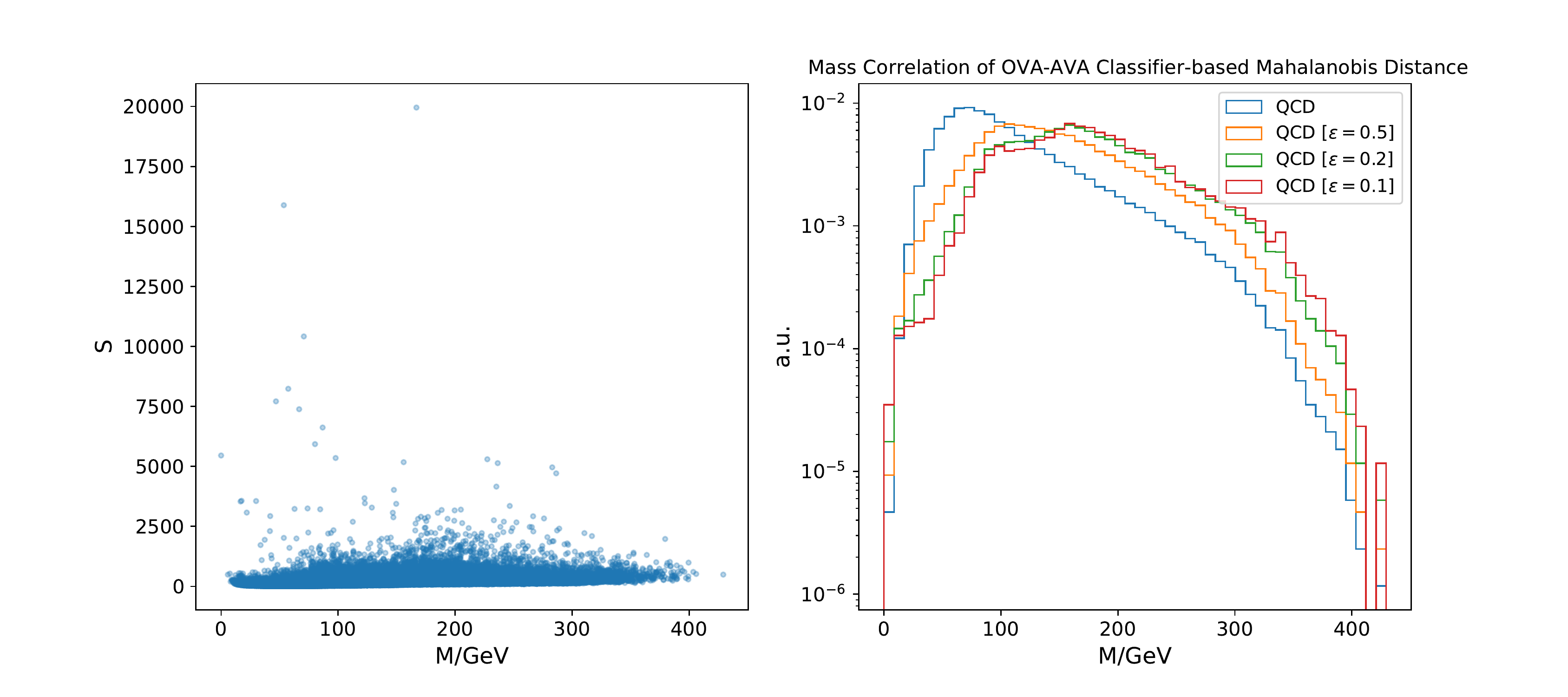}
    \includegraphics[width=\textwidth]{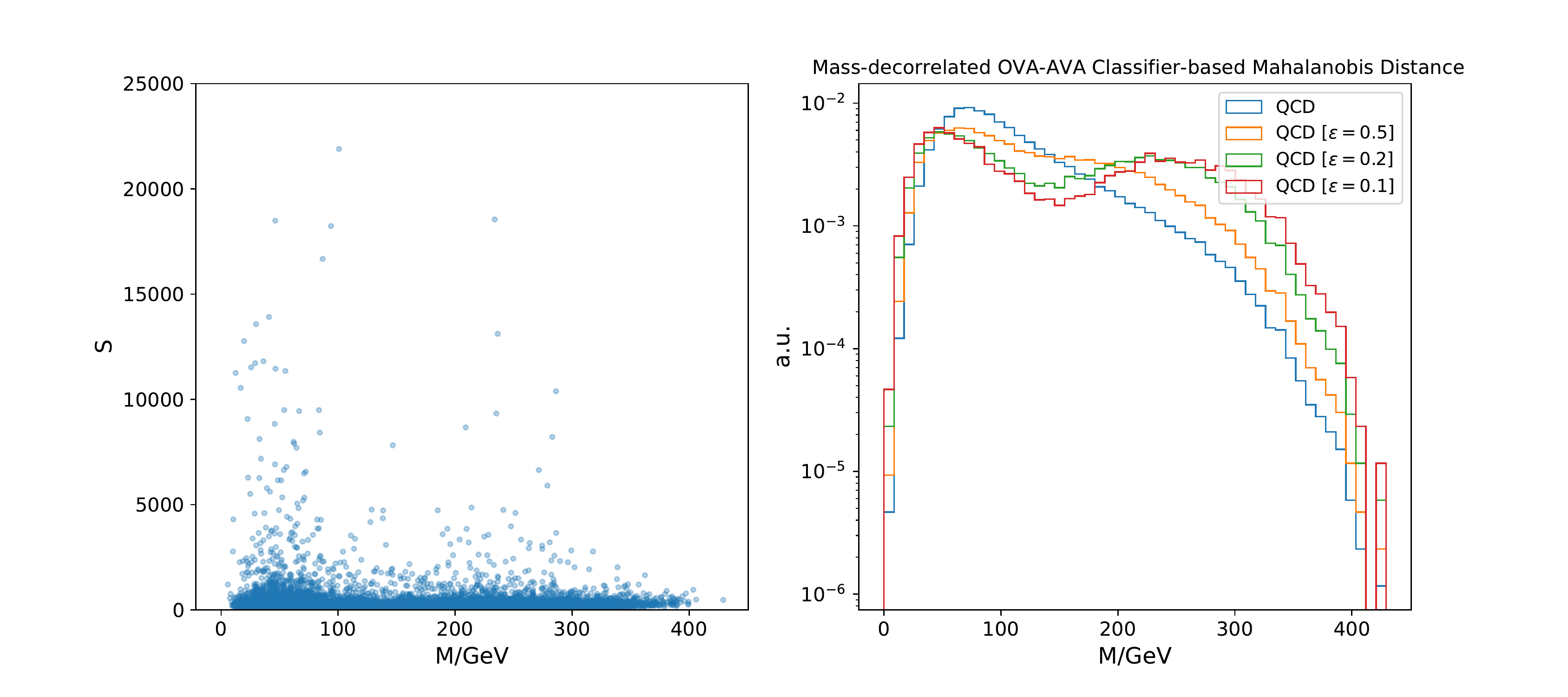}
        \caption{OVA-AVA Model.
            \emph{Top}: The un-decorrelated model with the Mahalanobis distance as the anomaly score.
            \emph{Bottom}: The mass-decorrelated model with the Mahalanobis distance as the anomaly score.
            \emph{Left}: The correlation between jet mass M and the anomaly score.
            \emph{Right}: Mass distributions for different background acceptance rates~($\epsilon$).\label{fig:mdeco_ovnni}}
\end{figure}

\begin{figure}
    \centering
    \includegraphics[width=\textwidth]{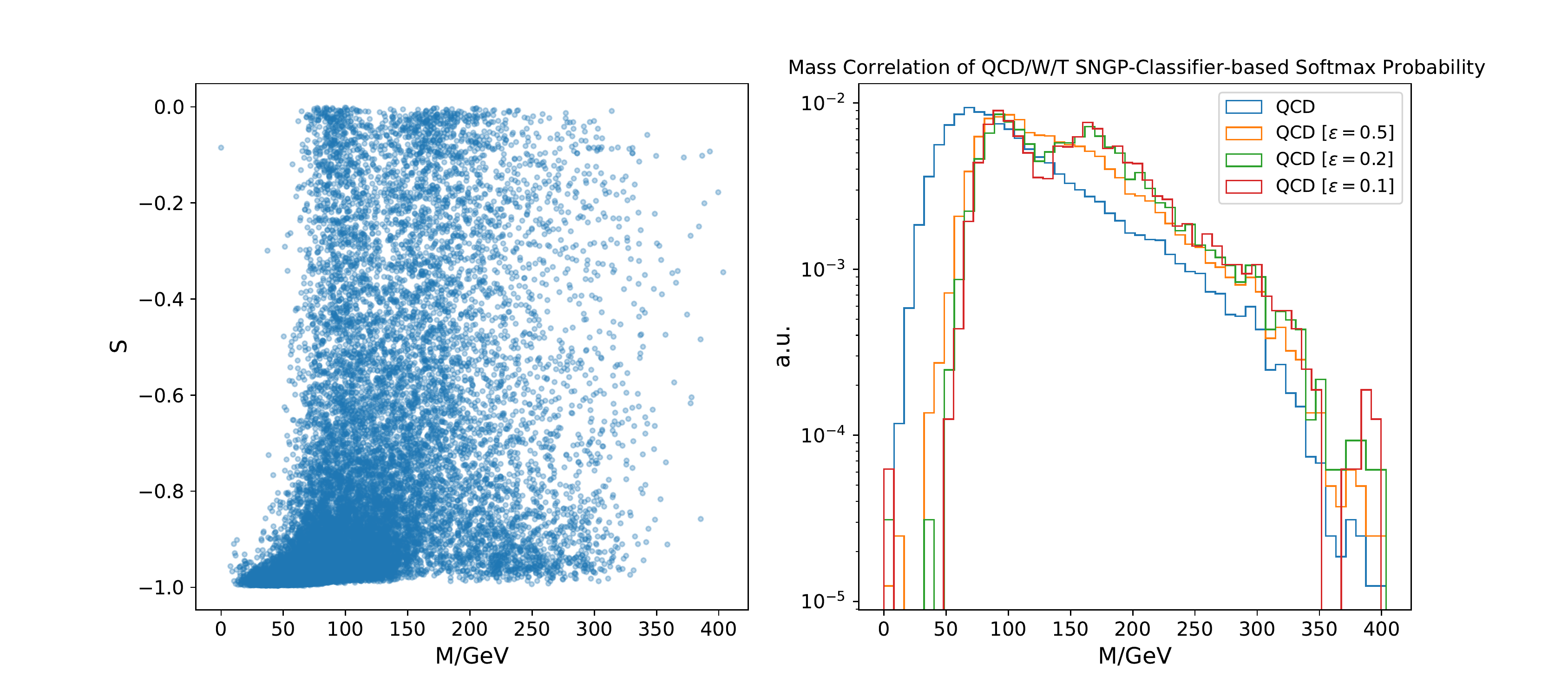}
    \includegraphics[width=\textwidth]{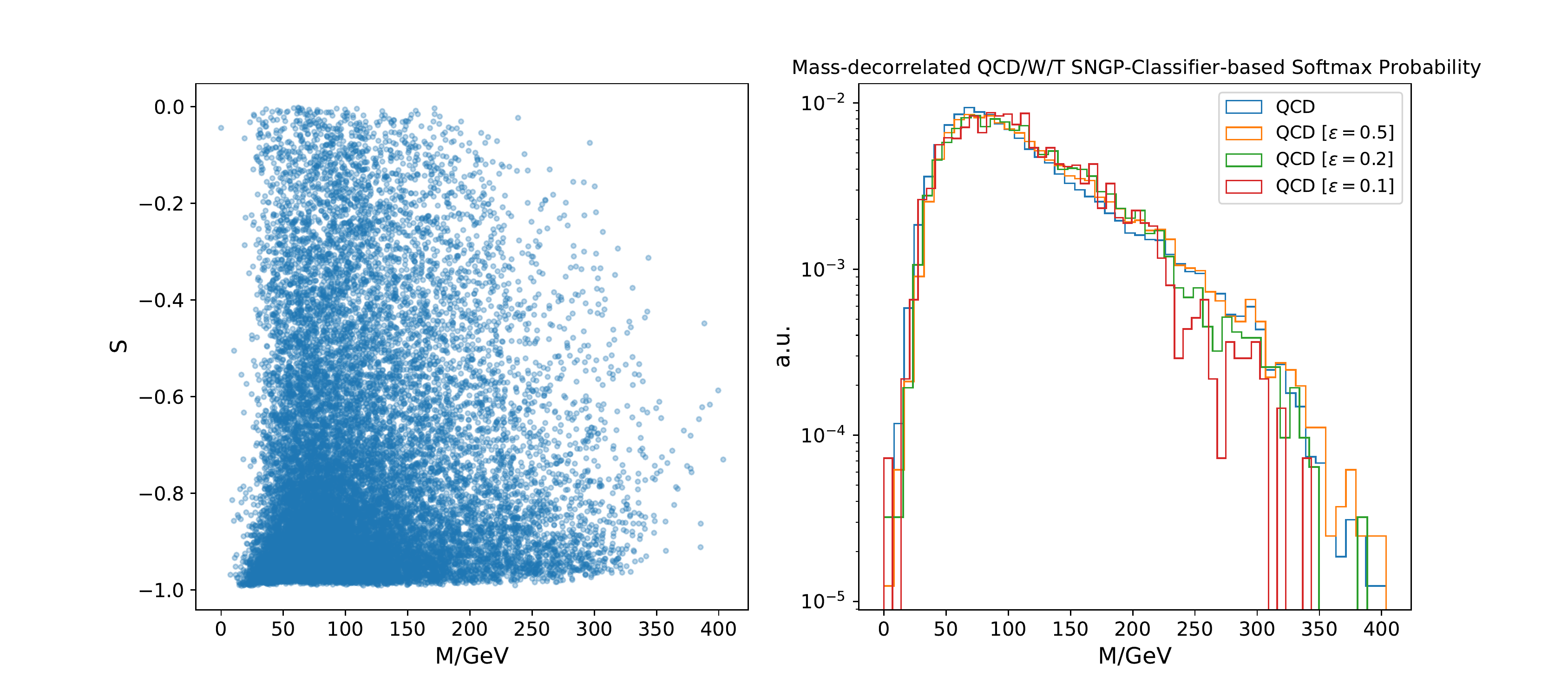}
    \caption{SNGP Model.
            \emph{Top}: The un-decorrelated model with $p_{\rm QCD}$ as the anomaly score.
            \emph{Bottom}: The mass-decorrelated model with $p_{\rm QCD}$ as the anomaly score.
            \emph{Left}: The correlation between jet mass M and the anomaly score.
            \emph{Right}: Mass distributions for different background acceptance rates ($\epsilon$).\label{fig:mcorr_sngp}}
\end{figure}

\begin{figure}
    \centering
    \includegraphics[width=\textwidth]{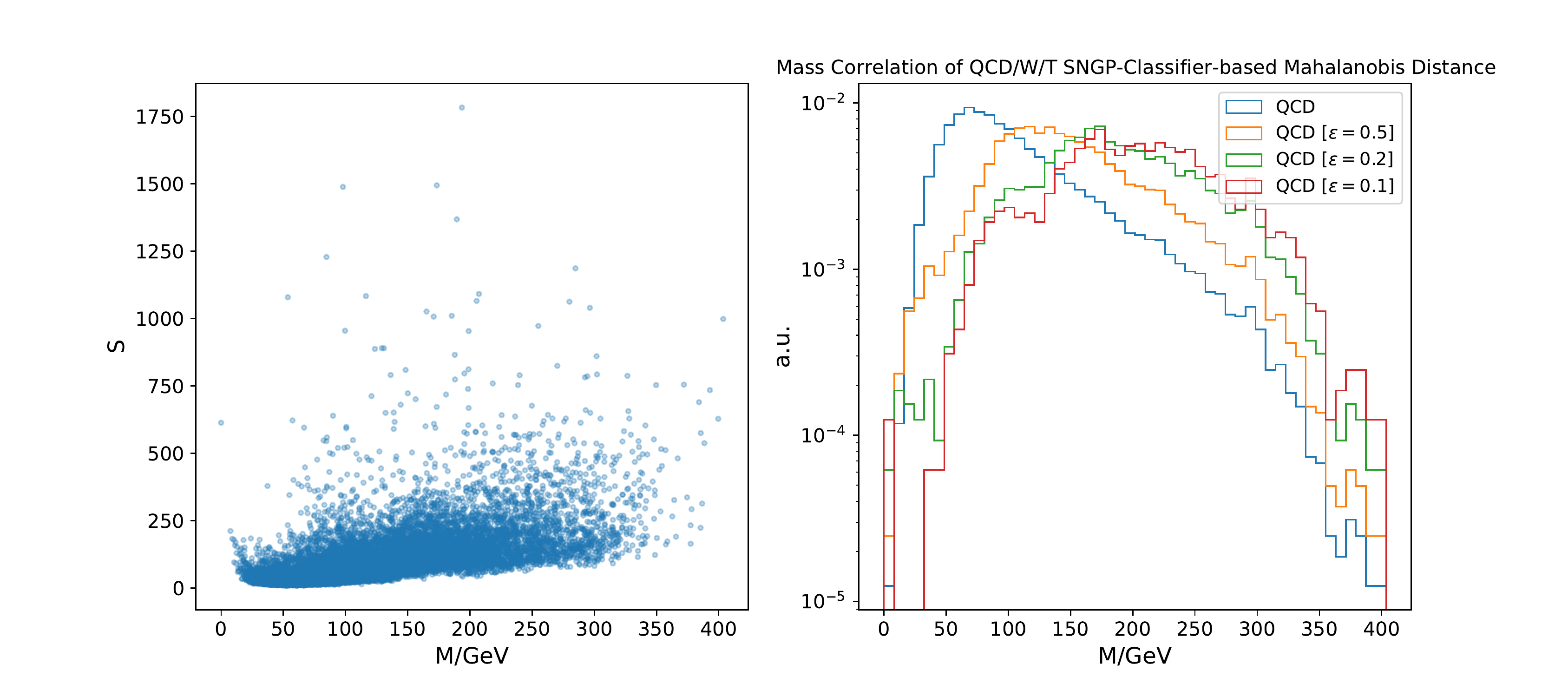}
    \includegraphics[width=\textwidth]{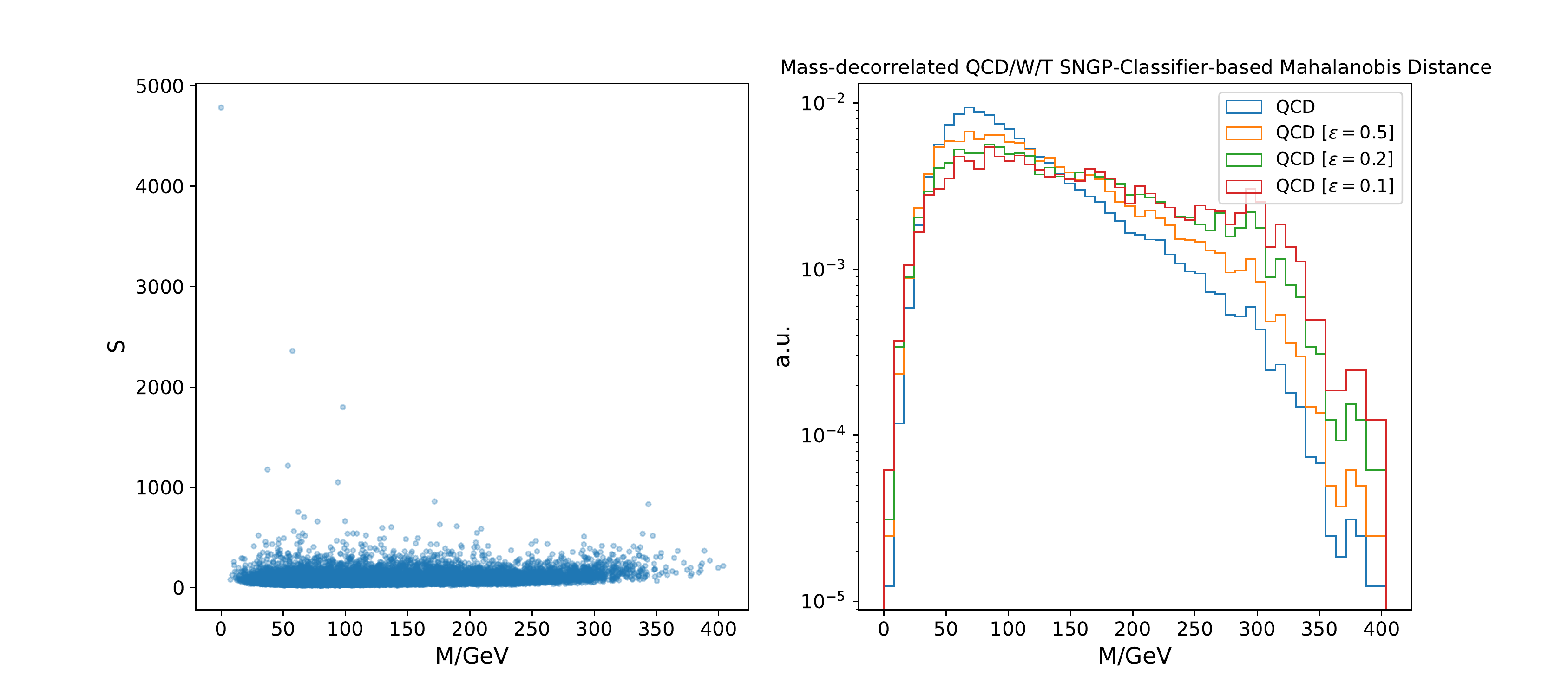}
    \caption{SNGP Model.
            \emph{Top}: The un-decorrelated model with the Mahalanobis distance as the anomaly score.
            \emph{Bottom}: The mass-decorrelated model with the Mahalanobis distance as the anomaly score.
            \emph{Left}: The correlation between jet mass M and the anomaly score.
            \emph{Right}: Mass distributions for different background acceptance rates ($\epsilon$).\label{fig:mdeco_sngp}}
\end{figure}

\begin{figure}
    \centering
    \includegraphics[width=\textwidth]{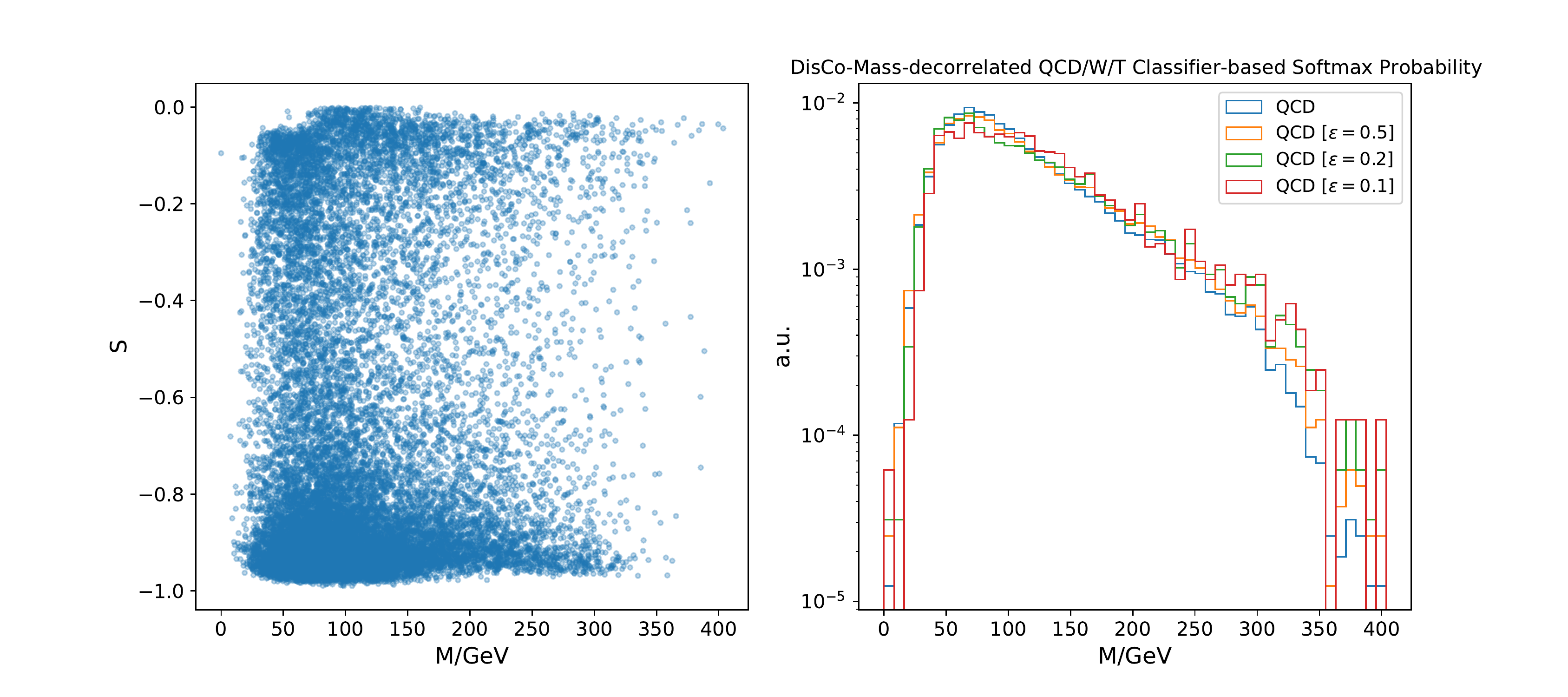}
    \includegraphics[width=\textwidth]{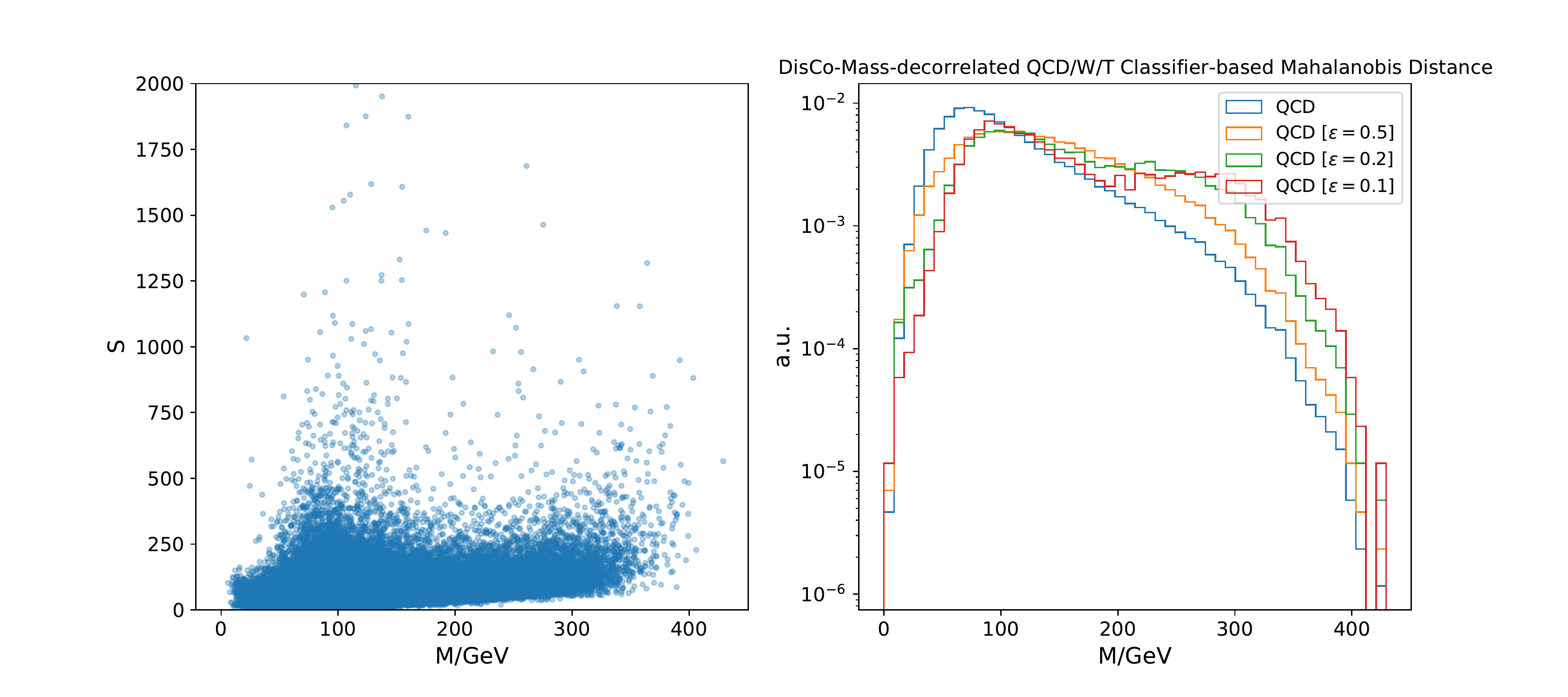}
        \caption{DisCo($\lambda=100$) Model.
            \emph{Top}: The mass-decorrelated model with $p_{\rm QCD}$ as the anomaly score.
            \emph{Bottom}: The mass-decorrelated model with the Mahalanobis distance as the anomaly score.
            \emph{Left}: The correlation between jet mass M and the anomaly score.
            \emph{Right}: Mass distributions for different background acceptance rates~($\epsilon$).\label{fig:mdeco_disco}}
\end{figure}

Figure~\ref{fig:mcorr_ensem},~\ref{fig:mdeco_ensem} show the mass (de)correlation results for the Ensemble model.
Figure~\ref{fig:mcorr_ovnni},~\ref{fig:mdeco_ovnni} and figure~\ref{fig:mcorr_sngp},~\ref{fig:mdeco_sngp} show the results for the OVA-AVA and the SNGP model respectively.
And in figure~\ref{fig:mdeco_disco}, we present the mass decorrelation results of the DisCo model.

\end{document}